\documentclass[final,5p,times,twocolumn]{elsarticle}
 \biboptions{comma,sort&compress}
\usepackage{graphicx}
\usepackage{here}
\usepackage[dvips]{epsfig}
\def\beq{\begin{equation}}
\def\eeq{\end{equation}}
\def\bea{\begin{eqnarray}}
\def\eea{\end{eqnarray}}
\def\bq{\begin{quote}}
\def\eq{\end{quote}}

\def\nnb{\nonumber}
\def\ga{\left(}
\def\dr{\right)}

\def\lrar{\Longrightarrow}
		
\def\nnb{\nonumber}
\def\la{\langle}
\def\ra{\rangle}
\def\nin{\noindent}
\def\ba{\vspace*{-0.2cm}\begin{array}}
\def\ea{\end{array}\vspace*{-0.2cm}}

\def\b{$\bullet~$}
\def\als{\alpha_s}

\def\gg2{ \la\alpha_s G^2 \ra}
\def\gg3{g^3f_{abc}\la G^aG^bG^c \ra}
\def\ggg4{\la\als^2G^4\ra}

\def\beq{\begin{equation}}
\def\enq{\end{equation}}
\def\beqa{\begin{eqnarray}}
\def\enqa{\end{eqnarray}}
\def\nnb{\nonumber}

\def\lb{\label}

\newcommand{\rag}{\rangle}
\newcommand{\lag}{\langle}

\journal{Elsevier}

\begin{document}

\begin{frontmatter}

\title{Improved light quark masses from pseudoscalar sum rules 
}
 \author[label2]{Stephan Narison}
\address[label2]{Laboratoire
Univers et Particules de Montpellier, CNRS-IN2P3, 
Case 070, Place Eug\`ene
Bataillon, 34095 - Montpellier, France.}
   \ead{snarison@yahoo.fr}
\begin{abstract}
\nin
Using ratios of the inverse Laplace transform sum rules within stability criteria for the subtraction point $\mu$ in addition to the ones of the usual $\tau$ spectral sum rule variable and continuum threshold $t_c$, we extract   the $\pi(1300)$ and $K(1460)$ decay constants to order $\alpha_s^4$ of perturbative QCD by including power corrections up to dimension-six condensates, tachyonic gluon mass,
instanton and finite width corrections. 
Using  these inputs with enlarged generous errors, we extract, {\it in a model-independent and conservative ways}, the sum of the scale-independent renormalization group invariant (RGI) quark masses $(\hat m_u+\hat m_q):q\equiv d,s$
and the corresponding running masses $(\overline{m}_u+\overline{m}_q)$ evaluated at 2 GeV. By giving the value of the ratio $m_u/m_d$, we deduce the running quark masses $\overline{m}_{u,d,s}$ and condensate $\la\overline{ \bar uu}\ra$ and the scale-independent mass ratios : $2m_s/(m_u+m_d)$ and $m_s/m_d$. Using the positivity of the QCD continuum contribution to the spectral function, we also deduce, from the inverse Laplace transform sum rules, for the first time to order $\alpha_s^4$, new lower bounds on the RGI masses which are translated into the running masses at 2 GeV and into upper bounds on the running quark condensate $\la\overline{ \bar uu}\ra$. Our results summarized in Table \ref{tab:res} and compared with our previous results and with recent lattice averages suggest that precise phenomenological determinations of the sum of light quark masses require improved experimental measurements of the $\pi(1.3)$ and $K(1.46)$ hadronic widths and/or decay constants which are the dominant sources of errors in the analysis.
\end{abstract}
\end{frontmatter}
\section{Introduction and a short historical overview}
Pseudoscalar sum rules have been introduced for the first time in \cite{BECCHI} for giving 
a bound on the sum of running light quark masses defined properly for the first time in the $\overline{MS}$-scheme  by \cite{FNR}.  Its Laplace transform version including power corrections introduced by SVZ \cite{SVZ} \footnote{For review, see e.g. \cite{RRY,SNB0,SNB1,SNB2,SNB3}.},\footnote{Radiative corrections to the exponential sum rules have been first derived in \cite{SNRAF}, where it has been noticed that the PT series has the property of an Inverse Laplace transform.} has been applied few months later to the pseudoscalar channel in \cite{SNRAF} and extended to the estimate of the SU(3) corrections to kaon PCAC in \cite{SN81}.  Its first application to the scalar channel was in \cite{SCAL}. Later on, the previous analysis has been reconsidered in \cite{TREL} for extracting e.g. the $\pi(1300)$ and $K(1460)$ decay constants. The first FESR analysis in the pseudoscalar channel has been done in \cite{LARIN,PSEU} which has been used later on by various authors \footnote{For reviews, see e.g.: \cite{SNB1,SNB2}.}. 

However, the light pseudoscalar channel is  {\it quite delicate} as the PT radiative corrections   (\cite{BECCHI,BROAD1} for the $\alpha_s$, \cite{LARIN,LARIN2} for the $\alpha_s^2$, \cite{CHET3} for the $\alpha_s^3$ and \cite{CHET4} for the $\alpha_s^4$ corrections) are quite large for low values of $Q^2\approx 1$ GeV$^2$ where the Goldstone pion contribution is expected to dominate the spectral function, while (less controlled) and controversial instanton-like contributions \cite{SHURYAK,IOFFE,NASON}\,\footnote{However, analogous contribution might lead to some contradiction in the scalar channel \cite{ZAKI}.}   might break the operator product expansion (OPE) at a such low scale. However, working at higher values of $Q^2$ for avoiding these QCD series convergence problems, one has to face  the dominant contribution from radial excited states where a little experimental information is known. Some models have been proposed in the literature for parametrizing the high-energy part of the spectral function. It has been proposed in \cite{TREL} to extract the $\pi$(1300) and K(1460) decay constants by combining the pesudoscalar and scalar sum rules which will be used in the Laplace sum rules for extracting the light quark masses. Though interesting,  the analysis was quite qualitative (no estimate of the errors) such that it is not competitive for an accurate determination of the quark masses. This estimate has been improved  in \cite{SNB1} using a narrow width approximation (NWA). Later on, a much more involved  ChPT based parametrization of the pion spectral function has been proposed in \cite{BPR} where the model dependence mainly appears in the interference between the $\pi(1300)$ and the $\pi(1800)$. Using FESR with some weight functions inspired from $\tau$-decay \cite{BNP2,BNP,SNTAU95,SNGTAU95,SNGTAU5,SNTAU9},
the authors of \cite{MALT} have extracted the decay constants of the $\pi(1300)$ and the $\pi(1800)$ by assuming that they do not interfere in the spectral function. The results for the spectral function are one of the main ingredient for extracting the light quark masses from pseudoscalar sum rules and it is important to have a good control (and a model-independence) of its value for a more precise and model-independent determination  of such light quark masses. 

In this paper, our aim is to extract the spectral function or the $\pi(1300)$ and K(1460)  decay constants from the ratio of Laplace sum rules known to order $\alpha_s^4$ of perturbation theory (PT) and  including power corrections up to dimension six within the SVZ expansion plus those beyond it such as the tachyonic gluon mass and the instanton contributions.  With this result, we shall extract the light quark mass values at the same approximation of the QCD series.
\section{The pseudoscalar Laplace sum rule}
 \subsection*{\b The form of the sum rules}
 \nin
We shall be concerned with the two-point correlator :
\beq
\psi^{P}_{5}(q^2)=i\int d^4x ~e^{iq\cdot x}\lag 0
|TJ^P_{5}(x)J^P_{5}(0)^\dagger
|0\rag~,
\lb{2po}
\eeq
where $J^P_{5}(x)$ is the local pseudoscalar current :
\beq
J^P_{5}(x)\equiv (m_u+m_q)\bar u(i\gamma_5)q~,~~~~q=d,~s~; ~P=\pi,~K~.
\label{eq:current}
\eeq
The associated Laplace sum rules (LSR)  ${\cal L}^P_{5}(\tau)$ and
its ratio ${\cal R}^P_{5}(\tau)$ read\,\cite{SVZ}\,\footnote{A quantum mechanics interpretation of these Laplace sum rules has been given by \cite{BELL}.}:
\beq
{\cal L}_5^{P}(\tau,\mu)=\int_{(m_u+m_q)^2}^{t_c}dt~e^{-t\tau}\frac{1}{\pi} \mbox{Im}\psi_5^P(t,\mu)~,
\label{eq:lsr}
\eeq
\beq\label{eq:ratiolsr}
{\cal R}^{P}_5 (\tau,\mu) = \frac{\int_{(m_u+m_q)^2}^{t_c} dt~t~ e^{-t\tau}\frac{1}{\pi}\mbox{Im}\psi^P_5(t,\mu)}
{\int_{(m_u+m_q)^2}^{t_c} dt~ e^{-t\tau} \frac{1}{\pi} \mbox{Im}\psi_5^{P}(t,\mu)}~,
\eeq
where $\mu$ is the subtraction point which appears in the approximate QCD series. 
 The ratio of sum  rules ${\cal R}_{5}^P (\tau,\mu)$ is useful here for extracting the contribution of the radial excitation $P'$ to the spectral function, while the Laplace sum rule ${\cal L}_5^{P}(\tau,\mu)$ will be used for determining the sum of light quark masses. 
  \subsection*{\b The QCD expression within the SVZ expansion}
  \nin
As mentioned earlier, the perturbative expression of the two-point correlator  $\psi^P_{5}(q^2)$ is known up to order $\alpha_s^4$ from successive works \cite{BECCHI,LARIN,LARIN2,CHET3,CHET4}. For a convenience of the reader, we give below the numerical expression\,\footnote{In the following, we shall not expand the QCD expression: $1/(1+ka_s+...+(np)\tau+...)$ as: $1-ka_s+k^2a_s^2+...-(np)\tau+(np)^2\tau^2+...$, but keep its non-expanded form.}:
\bea
{\cal L}_5^P(\tau)&=&{3\over 8\pi^2}(\overline{m}_u+\overline{m}_q)^2\tau^{-2}\Bigg{[} 1+\sum_{n=1,4} \delta^{(0)}_n a_s^n-\nnb\\
&&2{m_q^2\tau}\ga 1+\sum_{n=1,2}
\delta^{(2)}_n a_s^n\dr +\nnb\\
&&\tau^2 \delta^{(4)}+\tau^3\delta^{(6)}\Bigg{]}~,
\label{eq:lsrqcd}
\eea
where $\overline{m}_q$ is the running quark mass evaluated at the scale $\mu$. 
 From the analytic expression compiled in \cite{CHETL}, we derive the numerical PT corrections:
\bea
\delta^{(0)}_1&=&4.82107-2l_\mu~,\nnb\\
\delta^{(0)}_2&=&21.976-28.0729l_\mu+{17\over 4}l_\mu^2~,\nnb\\
\delta^{(0)}_3&=&53.1386-677.987l_\mu+102.82l_\mu^2-{221\over 24}l_\mu^3~,\nnb\\
\delta^{(0)}_4&=&-31.6283+756701l_\mu+1231.57l_\mu^2-\nnb\\
&& 321.968l_\mu^3+{7735\over 384}l_\mu^4~,\nnb\\
\delta^{(2)}_1&=&7.64213-4l_\mu~,\nnb\\
\delta^{(2)}_2&=&51.0915-62.93l_\mu+{25\over 2}l_\mu^2~,
\eea
where : $a_s\equiv \alpha_s/\pi$; $l_\mu\equiv -{\rm Log}(\tau\mu^2)$. The non-perturbative corrections are combinations of RGI quantities defined in \cite{SNT,SC88,BNP}:
\bea
\overline {m_q\la \bar qq\ra}&=&m_q\la \bar qq\ra+{3\over 7\pi^2}m_q^4\ga {1\over a_s}-{53\over 24}\dr\nnb\\
\overline{\la \alpha_sG^2\ra}&=&\la \alpha_s G^2\ra \ga 1+{16\over 9}a_s\dr 
-{16\over 9}\alpha_s\ga 1+{91\over 24}a_s\dr m_q\la \bar qq\ra~, \nnb\\
&&-{1\over 3\pi}\ga 1+{4\over 3}a_s\dr m_q^4 ~.
\eea
In terms of these quantities, they read \cite{BECCHI,SNB1,BROAD,JM95}:
\bea
\delta^{(4)}={4\pi^2\over 3}\ga \delta^{(4)}_q+\delta^{(4)}_g\dr~,
\eea
with:
\bea
\delta^{(4)}_q&=&-2m_q\la \bar uu\ra \Big{[} 1+a_s\ga 5.821-2l_\mu\dr\Big{]}+\nnb\\
&&\overline {m_q\la \bar qq\ra}\Big{[} 1+a_s\ga 5.266--2l_\mu\dr\Big{]}-\nnb\\
&&{3\over 7\pi^2}m_q^4\ga {1\over a_s}+2.998-{15\over 4}l_\mu\dr~,\nnb\\
\delta^{(4)}_g&=&{1\over 4\pi}\overline{\la \alpha_sG^2\ra}\Big{[} 1+a_s\ga 4.877-2l_\mu\dr\Big{]}~.
\eea
The contribution of the $d=6$ condensate is:
\bea
\delta^{(6)}&=&-{4\pi^2\over 3}\Big{[} m_q\la \bar uGu\ra +\nnb\\
&&{32\over 27}\pi \rho\alpha_s\ga \la\bar uu\ra^2+\la\bar qq\ra^2-9\la\bar uu\ra\la\bar qq\ra\dr\Big{]}~,
\eea
where $\la \bar uGu\ra\equiv \la \bar u(\lambda^a/2)G^{\mu\nu}_a\sigma_{\mu\nu}u\ra\equiv M^2_0
\la \bar uu\ra$ with $M^2_0=(0.8\pm 0.2)~\rm{GeV}^2$ \cite{JAMI2,HEID,SNhl} is the quark-gluon mixed condensate; $\rho
=(4.2\pm 1.3)$\,\cite{SNGTAU95,JAMI2,LNT}
indicates the violation of the vacuum saturation assumption of the four-quark operators.  

  \subsection*{\b Tachyonic gluon mass and estimate of larger order PT-terms}
  \nin
  The tachyonic gluon mass $\lambda$ of dimension two has been introduced in \cite{AKHOURY,CNZ} and appears naturally in most holographic QCD models\,\cite{JUGEAU}.
Its contribution is ``dual" to the uncalculated higher order terms of the PT series \cite{SNZ} and disappears
for long PT series like in the case of lattice calculations \cite{PINEDA}, but should remain when only few terms of the PT series are calculated like in the case studied here.  
Its contribution reads \cite{CNZ}:
  \beq
{\cal L}_5^P(\tau)\vert^{tach}=-{3\over 2\pi^2}(\overline{m}_u+\overline{m}_q)^2a_s\lambda^2\tau^{-1}~,
\eeq
Its value  has been estimated  from $e^+e^-$ \cite{SNGTAU95,SNe} and $\tau$-decay \cite{SNGTAU5} data:
\beq
a_s\lambda^2=-(0.07\pm 0.03)~{\rm GeV}^2~.
\eeq
  \subsection*{\b The instanton contribution}
  \nin
 The inclusion of this contribution into the operator product expansion (OPE) is not clear and controversial \cite{SHURYAK,IOFFE,NASON}. In addition, an analogous contribution might lead to some contradiction to the OPE in the scalar channel \cite{ZAKI}. Therefore, we shall consider  the sum rule including the instanton  contribution as an alternative approach. For our purpose, we parametrize this contribution as in \cite{SHURYAK,IOFFE}, where its  corresponding contribution to the Laplace sum rule reads:
\beq
{\cal L}_5^P(\tau)\vert^{inst}={3\over 8\pi^2}(m_u+m_q)^2\tau^{-3} \rho_c^2e^{-r_c}\big{[}K_0(r_c)+K_1(r_c)\big{]}~,
\label{eq:instanton}
\eeq
where $ K_i$ is the Bessel-Mac-Donald function; $r_c\equiv \rho_c^2/(2\tau)$ and $\rho_c=(1.89\pm 0.11) $ GeV$^{-1}$ \cite{SNH10} is the instanton radius.
  \subsection*{\b Duality violation }
  \nin
Some eventual additional contribution from duality violation (DV) \cite{SHIF} could also be considered. However, as the LSR use the OPE in the Euclidian region where the DV effect is exponentially suppressed, one may safely neglect such contribution in the present analysis \footnote{We thank the 2nd referee for this suggestion and for different provocative comments leading to the improvements of the final manuscript.}. 
{\scriptsize
\begin{table}[hbt]
\setlength{\tabcolsep}{0.9pc}
 \caption{
Input parameters: the value of $\hat\mu_q$ has
been obtained  from the running 
masses evaluated at 2 GeV:  $(\overline{m}_u+\overline{m}_d)=7.9(6)$ MeV \cite{SNB1,SNmass}. 
Some other predictions and related references can be found in \cite{PDG}; $\rho$ denotes the deviation on the estimate of the four-quark condensate from vacuum saturation. The error on
$\Gamma_{K'}$ is a guessed conservative estimate. 
Most of the original errors have been enlarged for a conservative estimate of the errors.    
 }
    {\small
\begin{tabular}{lll}
&\\
\hline
Parameters&Values& Ref.    \\
\hline
$\Lambda(n_f=3)$& $(353\pm 15)$ MeV &\cite{SNTAU9,BETHKE}\\
$\hat m_s$&$(0.114\pm0.021)$ GeV &\cite{SNB1,SNTAU9,SNmass,SNmass98}\\
$\hat \mu_d$&$(253\pm 6)$ MeV&\cite{SNmass,SNmass98}\\
$\kappa\equiv \la \bar ss\ra/\la\bar dd\ra$& $(0.74^{+0.34}_{- 0.12})$&\cite{HBARYON,SNB1}\\
$-a_s\lambda^2$& $(7\pm 3)\times 10^{-2}$ GeV$^2$&\cite{SNe,SNGTAU5}\\
$\la\alpha_s G^2\ra$& $(7.0\pm 2.6)\times 10^{-2}$ GeV$^4$&
\cite{SNH10}\\
$M_0^2$&$(0.8 \pm 0.2)$ GeV$^2$&\cite{JAMI2,HEID,SNhl}\\
$\rho \alpha_s\la \bar qq\ra^2$&$(5.8\pm 1.8)\times 10^{-4}$ GeV$^6$&\cite{SNTAU95,LNT,JAMI2}\\
$\rho_c$&$(1.89\pm 0.11) $ GeV$^{-1}$&\cite{SNH10}\\
$\Gamma_{\pi'}$&$(0.4\pm 0.2)$ GeV&\cite{PDG}\\
$\Gamma_{K'}$&$(0.25\pm 0.05)$ GeV&\cite{PDG}\\
\hline
\end{tabular}
}
\label{tab:param}
\end{table}
} 
  \subsection*{\b The  QCD input parameters}
  \nin
 There are several estimates of the QCD  input parameters in the current literature using different
approaches and sometimes disagree each others. For a for self-consistency, we shall work in this paper with the input parameters given in  Table {\ref{tab:param}} obtained using the same approach (Laplace or/and $\tau$-decay-like sum rule) as the one used here and wihin the same criterion of stability (minimum, maximum or plateau in $\tau$ and $t_c$).

-- $\hat m_q$ and $\hat\mu_q$ are RGI invariant mass and condensates which are related
to the corresponding running parameters as~\cite{FNR}:
\bea
{\overline m}_{q}(\tau)&=&
{\hat m}_{q}  \ga-\beta_1a_s\dr^{-2/{
\beta_1}}\ga 1+\rho_m\dr 
\nnb\\
{\la\overline{\bar qq}\ra}(\tau)&=&-{\hat \mu_q^3  \ga-\beta_1a_s\dr^{2/{
\beta_1}}}/\ga 1+\rho_m\dr 
\nnb\\
{\la\overline{\bar q Gq}\ra}(\tau)&=&-{M_0^2{\hat \mu_q^3} \ga-\beta_1a_s\dr^{1/{3\beta_1}}}/\ga 1+\rho_m\dr ~,
\label{eq:rgi}
\eea
where $\beta_1=-(1/2)(11-2n_f/3)$ is the first coefficient of the QCD $\beta$-function for $n_f$-flavours. $\rho_m$ is the QCD correction which reads to N4LO accuracy for $n_f=3$ \cite{SNB1,RUNDEC}:
\beq
\rho_m=0.8951a_s+1.3715a_s^2+0.1478a_s^3~,
\eeq
where $a_s\equiv \alpha_s/\pi$ is the QCD running coupling.
The value of $\hat m_s$ quoted in Table \ref{tab:param} will serve as an initial value for the $m_s$ corrections in the PT expression of the kaon correlator. It will be re-extracted by iteration in the estimate of $m_s$ from the kaon sum rule where one obtains a convergence of the obvious iteration procedure after two iterations. 

-- The value of the $\mu_q$ RGI condensate used in Table \ref{tab:param} comes from the value  $(\overline{m}_u+\overline{m}_d)=(7.9\pm 0.6)$ MeV evaluated at 2 GeV from \cite{SNmass} after the use of the GMOR relation:  
\beq
2m_\pi^2f_\pi^2=-(m_u+m_d)\la\bar uu+\bar dd\ra~,
\label{eq:gmor}
\eeq 
where $f_\pi=(92.23\pm 0.14)$ MeV \cite{ROSNER}. 

-- The value of the gluon condensate used here comes from recent charmonium sum rules. Since SVZ, several determinations of the gluon condensates exist in the literature \cite{IOFFE,SNGTAU95,LNT,SNe,fesr,YNDU,SNHeavy,BELL,NEUF,SNG,SNGTAU5}. The quoted error is about 2 times the original error for making this value compatible with the SVZ original value and charmonium analysis in \cite{IOFFE} commented in \cite{SNH10}\,\footnote{The sets of FESR in \cite{fesr} tend give large values of the condensates which are in conflict with the ones from LSR and $\tau$-like sum rules using similar $e^+e^-$ data  \cite{SNTAU95,SNGTAU95,LNT,SNe} and previous charmonium analysis \cite{SVZ,IOFFE,BELL,SNH10}. They will not be considered here. However, as shall see explicitly later on, the effects of $\la\alpha_s G^2\ra$ and of the tachyonic gluon mass used in this paper are relatively small in the present analysis.}.

-- We use the value of the four-quark condensate obtained from $e^+e^-$ and VV+AA $ \tau$-decay  \cite{SNTAU95,SNGTAU95,SNGTAU5,LNT}  data and from light baryons sum rules \cite{JAMI2} where a deviation from the vacuum saturation by a factor $\rho\simeq (4.2\pm 1.3)$ has been obtained if one evaluates $\la\bar dd\ra$ from $\mu_d$ given in Table \ref{tab:param} at $M_\tau$ where the four-quark condensate has been extracted (for a conservative result, we have multiplied the original error in \cite{SNTAU95} by a factor 2). Similar conclusions have been derived from FESR \cite{fesr} and more recently from the VV-AA component of $\tau$-decay data \cite{SNG,BOITO}. We assume that a similar deviation holds in the pseudoscalar channels. We shall see again later on that the error induced by this contribution on your estimate is relatively negligible. 

-- We use the value of the SU(3) breaking parameter $\kappa\equiv \la \bar ss\ra/\la\bar dd\ra$ from \cite{HBARYON} which agrees with the ones obtained from from light baryons \cite{JAMI2} and from kaon and scalar \cite{SN81,TREL,SNB2} sum rules recently reviewed in \cite{SNB1} but
more accurate. For a conservative estimate we have 
enlarged the original error  by a generous factor 4 and the upper value for recovering 
the central value $\kappa=1.08$  from recent lattice calculations \cite{UKQCD}.
\subsection*{\b LSR $\tau$, $t_c$ and $\mu$ stability criteria} 
\nin
The LSR is obtained within approximation both for the spectral side (when data are not available 
like here) and for the QCD side (as one has to truncate the PT series and the OPE at given orders). 
In the ideal case, where both sides of the LSR are perfectly described, one expects to find a large range of plateau stability (exact matching) at which one can extract the resonance parameters. It often happens that the minimal duality ansatz: ``one resonance +QCD continuum from a threshold $t_c$" description of the spectral function and/or the QCD approximation is rather crude. In this case, one can still extract an optimal information on the resonance parameters if the curves present a minimum, maximum or inflexion point versus the external LSR variable $\tau$ and the continuum threshold $t_c$ as demonstrated in series of papers by Bell-Bertlmann \cite{BELL} using the examples of harmonic oscillator and non-relativistic charmonium sum rules (see e.g.  Figs. 49.6 and 49.7 pages 511-512 in Ref. \cite{SNB1}).  At this minimum, maximum or inflexion point, one has a narrow sum rule window at which there is  a balance between the QCD continuum and NP contributions  and where the OPE still makes sense and the lowest resonances relatively dominates the spectral function. Analysis based on these criteria have been applied successfully in different applications of the sum rules  (see e.g. Refs. \cite{SNB1,SNB2}). To these well-known $\tau$ and $t_c$ stability criteria, 
we require a $\mu$-stability  of the results in order to limit the arbitrariness of the subtraction point $\mu$ often chosen ad hoc in the existing literature. 
Throughout this paper, we shall use the above criteria for extracting the optimal results from the analysis.
\section{A Laplace sum rule estimate of the decay constant $f_{\pi'}$}
  \subsection*{\b The spectral function}
  \nin
  We shall parametrize the spectral function as:
\beq
\frac{1}{\pi}\mbox{ Im}\psi^\pi_{5}(t)\simeq \sum_{\pi,\pi'}2f^2_{P}m^4_{P}\delta(t-m_{P}^2)
  \ + \
 \mbox{QCD cont.} \theta (t-t_c),
\label{eq:duality}
\eeq
where the higher states ($\pi''$,...) contributions  are smeared by the ``QCD continuum" coming from the discontinuity of the QCD diagrams and starting from a constant threshold $t_c$.  $f_P$ is the well-known decay constant :
\beq
\la 0 \vert J^P_{5}(x) \vert P\ra =\sqrt{2} m_P^2f_P~,
\eeq
normalized here as: $f_\pi=(92.23\pm 0.14)$ MeV and $f_K\simeq (1.20\pm 0.01) f_\pi$ \cite{ROSNER}. 
We improve the $\pi'\equiv \pi(1300)$ contribution by taking into account the finite width correction by replacing the delta function with a Breit-Wigner shape:
\beq
\pi\delta(t-m_{\pi'}^2)\to BW(t)= {m_{\pi'}\Gamma_{\pi'}\over (t-m_{\pi'}^2)^2+m^2_{\pi'}\Gamma^2_{\pi'}}.
\label{eq:bw}
\eeq
Defined in this way, the $\pi'$ can be considered as an ``effective resonance" parametrizing the higher state contributions not smeared by the QCD continuum and may take into account some possible interference between the $\pi(1300)$ and $\pi(1800)$ contributions.
\subsection*{\b $f_{\pi'}$ from ${\cal R}^{\pi}_5$  within the SVZ expansion at arbitrary $\mu$\,\footnote{Here and in the following we shall denote by SVZ expansion the OPE without the instanton contribution.}}
  \nin
 One expects from some chiral symmetry arguments that $f_\pi'$ behaves like $m_\pi^2$. Therefore, one may expect that the $\pi'$ will dominate over the pion contribution in the derivative of the Laplace sum rule:
 \beq
 -{\partial \over \partial \tau}{\cal L}_5^\pi(\tau,\mu)~,
\eeq
from which one can extract the decay constant $f_\pi'$ or the $\pi(1300)$ contribution to the spectral function. In order to eliminate the unknown value of the sum of light quark masses $(m_u+m_d)$, it is convenient to work with the ratio of Finite Energy Laplace sum rules $
{\cal R}_5^\pi(\tau,\mu) $ defined in Eq. (\ref{eq:ratiolsr}). In so doing, we define the quantity:
\beq
r_\pi\equiv {M^4_{\pi'}f^2_{\pi'}\over m_\pi^4 f_\pi^2}~,
\eeq
which quantifies the relative weight between the $\pi'$ and the pion contribution into the spectral function. It is easy to deduce the sum rule:
\beq
r_\pi={{\cal R}^{\pi}_5 \vert_{qcd}-m_\pi^2\over BWI_1-{\cal R}^{\pi}_5 \vert_{qcd}BWI_0}e^{-m_\pi^2\tau}~.
\eeq
${\cal R}^{\pi}_5 \vert_{qcd}$ is the QCD expression of the FESR in Eq. (\ref{eq:ratiolsr}) where we have parametrised the spectral function by a step function corresponding to the perturbative expression for massless quarks from the threshold  $t_c$. $BWI_n$ is the Breit-Wigner integral:
\beq
BWI_n\equiv {1\over\pi}\int_{9m_\pi^2}^{t_c}dt~t^ne^{t\tau} BW(t)~:~~~~~n=0,1~,
\label{eq:bwi}
\eeq
where $BW(t)$ has been defined in Eq. (\ref{eq:bw}).

\begin{figure}[hbt] 
\begin{center}
\centerline {\hspace*{-8.5cm} a) }\vspace{-0.6cm}
{\includegraphics[width=8cm  ]{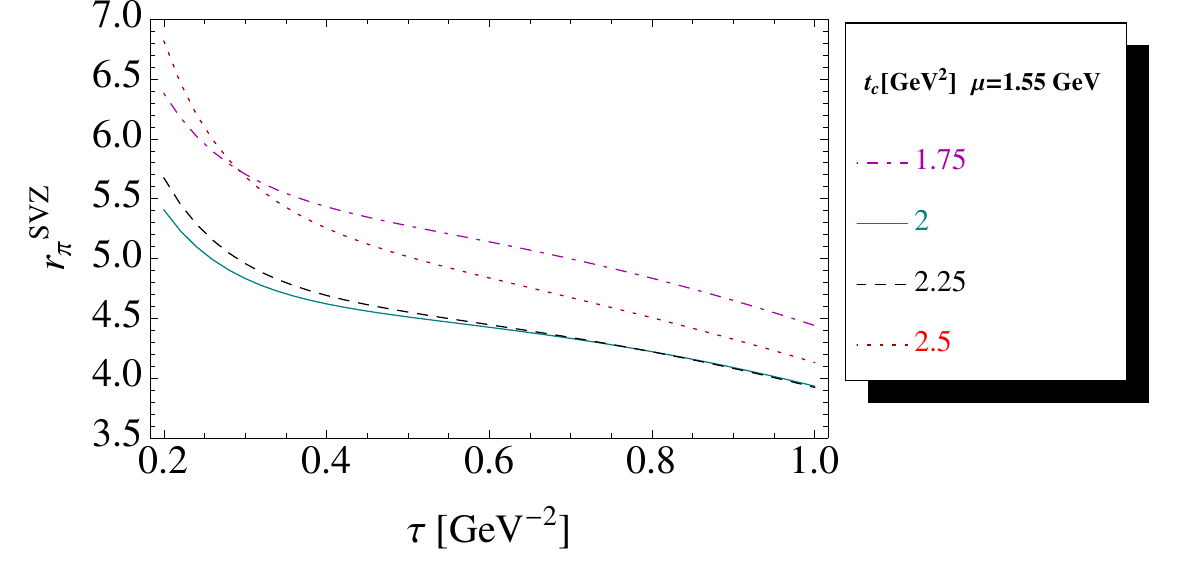}}
\centerline {\hspace*{-8.5cm} b) }\vspace{-0.3cm}
{\includegraphics[width=7cm  ]{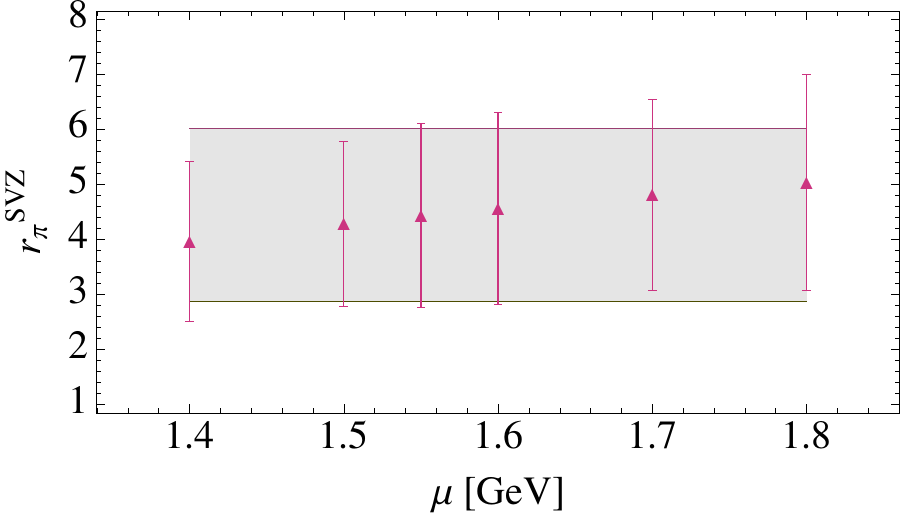}}
\centerline {\hspace*{-8.5cm} c) }\vspace{-0.3cm}
{\includegraphics[width=9cm  ]{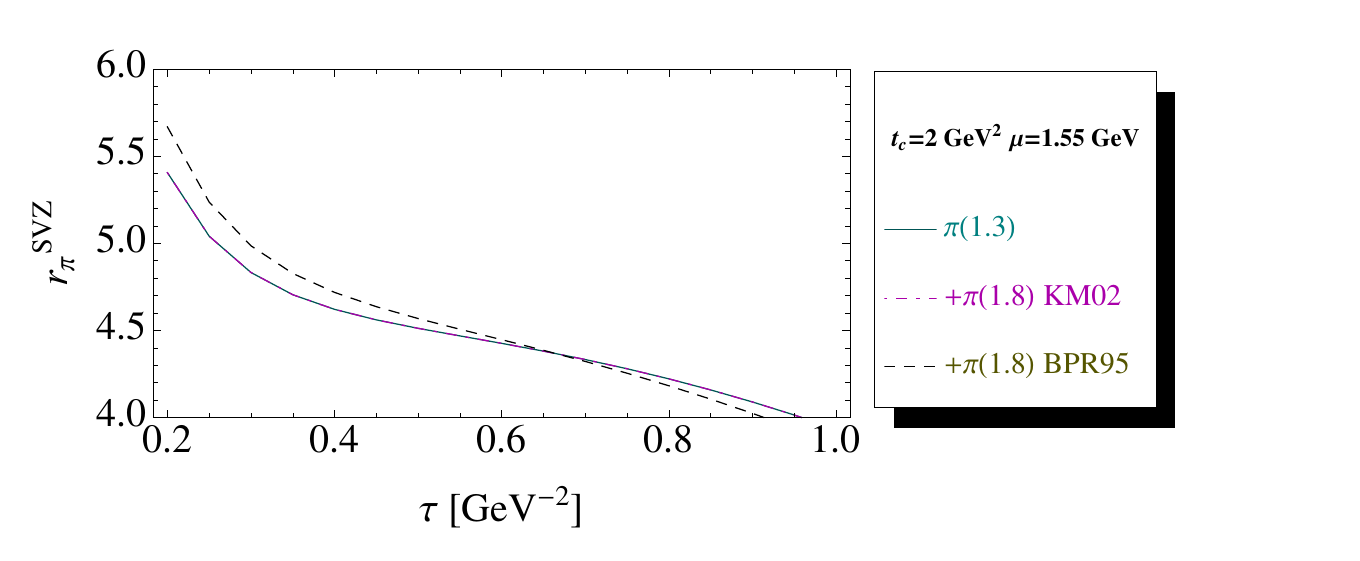}}
\caption{
\scriptsize 
{\bf a)} $\tau$-behaviour  of  $r_\pi$ for $\mu=1.55$ GeV  and for different values of $t_c$
within the SVZ expansion. 
; {\bf b)} $\mu$-behaviour  of  the optimal value of $r_\pi$ deduced from a). The coloured region corresponds to the mean value where the error comes from 
the most precise determination $\oplus$ the same systematics; {\bf c)} Comparison of the effects of $\pi(1.8)$ for two different models of the spectral function where one can remark a complete co\"\i ncidence of the $\pi(1.3)$ curve and $\pi(1.3)+\pi(1.8)$ 
from KM02. 
}
\label{fig:rpi}
\end{center}
\end{figure}

-- With the set of input parameters in Table \ref{tab:param}, we show in Fig. \ref{fig:rpi}a the $\tau$-behaviour of $r_\pi$ at a given value of $\mu=1.55$ GeV. We extract the optimal result at the value of $t_c=2$ GeV$^2$  and $\tau=(0.6\pm 0.1)$ GeV$^{-2}$ where both a minimum in the change of $t_c$ and an inflexion point in $\tau$ are obtained. One can notice that this value of $t_c$ slightly higher than the $\pi(1.3)$ mass is inside the region of best stability 
between the spectral function and the QCD expression studied explicitly in \cite{BPR}. For $\mu=1.55$ GeV, at which we have an inflexion point for the central value, we deduce:
\bea
r_\pi^{svz}&=&4.43(14)_{\Lambda}(4)_{\lambda^2}(13)_{\bar uu}(31)_{G^2}(1)_{\bar uGu}(20)_\rho\nnb\\
&&(161)_{\Gamma_{\pi}}(2)_{t_c}(10)_\tau\nnb\\
&=&4.43\pm 1.67~,
\label{eq:rpi_opea}
\eea
where the dominant error comes from the experimental width of the $\pi(1300)$ which needs to be improved. The errors due to the QCD parameters are negligible despite the enlarged errors introduced for a conservative result. 

-- In Fig. \ref{fig:rpi}b, we study the influence of the choice of $\mu$ varying in the range 1.4 to 1.8 GeV where a good duality between the QCD and spectral sides of the sum rules is obtained~ \cite{BPR}.  BPR has also noticed that  the value of $t_c (s_0=\mu^2$ in their notation) is below the $\pi(1.8)$ mass and there is a complex interference between the $\pi(1.3)$ and $\pi(1.8)$ indicating the complexity of the pseudoscalar spectral function. Therefore, for quantifying the $\pi(1.8)$ contribution, we study in Fig. \ref{fig:rpi}c, its effect by using one of the models proposed by BPR with the mixing parameter $\zeta=0.234+i~0.1$ which is the one which reproduces the best fit to the experimental curves which observe the $\pi(1.8)$ in hadronic interactions \cite{BPR}. We also compare our results with the one in Ref. \cite{MALT}
by taking the central value: $f_{\pi(1.8)}\approx  1.36$ MeV where no interference with the $\pi(1.3)$ has been assumed. We notice from the analysis in Fig. \ref{fig:rpi}c that, in both cases, the $\pi(1.8)$ effect is negligible. 

-- Our final result corresponds to the mean of different determinations in Fig. \ref{fig:rpi}b, where the dashed coloured region corresponds to  the final error $\pm 1.56$ from the most precise determination $\oplus$ the systematics 0.17 corresponding to the
distance of the mean to this precise determination. 
 added quadratically:
\beq
r_\pi^{svz}=4.45\pm1.56\pm 0.17_{syst}~,
\label{eq:rpi_ope}
\eeq
\nin
\begin{figure}[hbt] 
\begin{center}
\centerline {\hspace*{-8.5cm} a) }\vspace{-0.6cm}
{\includegraphics[width=7.5cm  ]{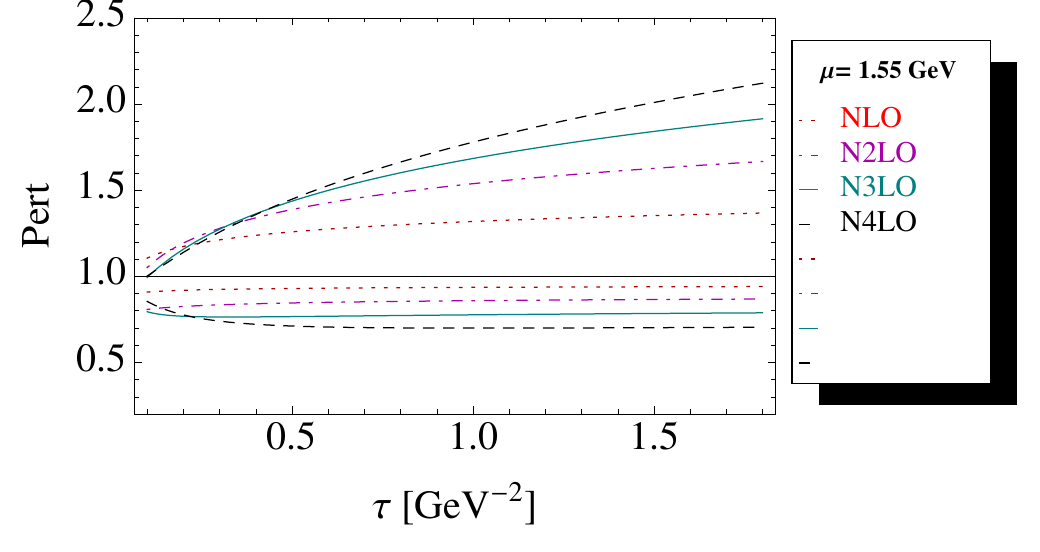}}
\centerline {\hspace*{-8.5cm} b) }\vspace{-0.3cm}
{\includegraphics[width=7.5cm  ]{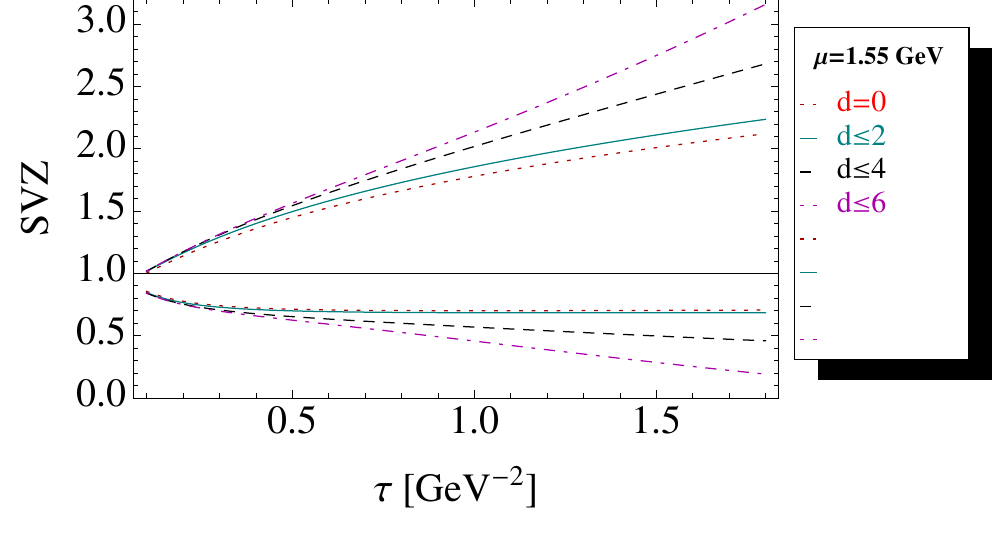}}
\caption{
\scriptsize 
{\bf a)} $\tau$-behaviour  of  the PT series of $\sqrt{\cal L}_5^{\pi}(\tau,\mu)$ (upper group of curves) and of ${\cal R}^{\pi}_5 (\tau,\mu)$ (lower group of curves)  appropriately normalised to 1 for $\tau=0$ and using $\mu=1.55$ GeV. 
; {\bf b)} the same as a) but for the power corrections within the SVZ expansion. 
}
\label{fig:conv}
\end{center}
\end{figure} 
\nin
\begin{figure}[hbt] 
\begin{center}
\centerline {\hspace*{-8.5cm} a) }\vspace{-0.6cm}
{\includegraphics[width=7.5cm  ]{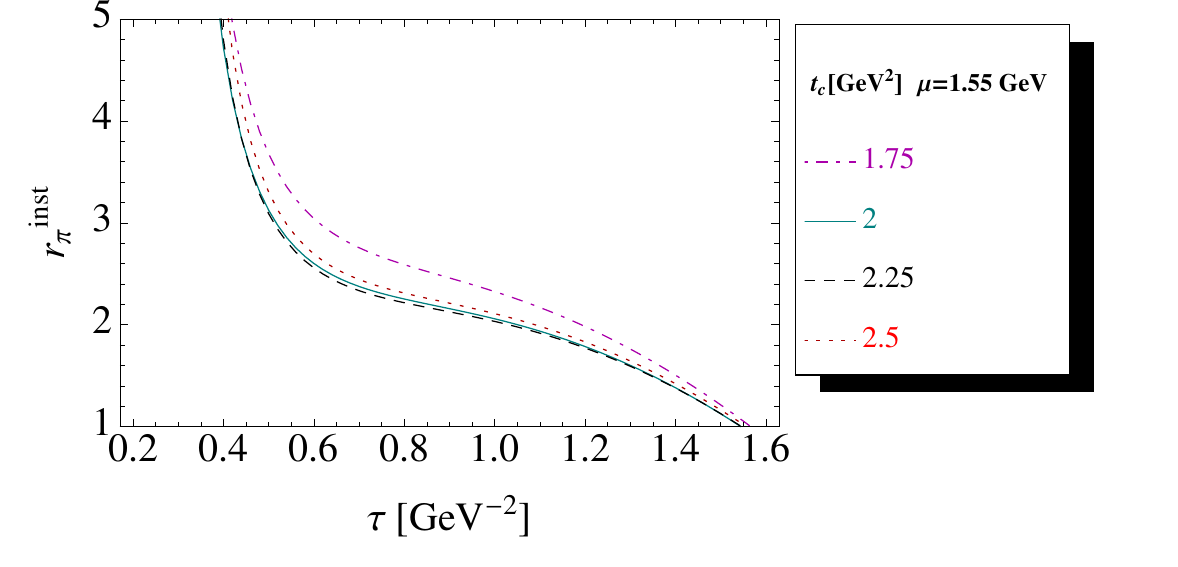}}
\centerline {\hspace*{-8.5cm} b) }\vspace{-0.3cm}
{\includegraphics[width=6.5cm  ]{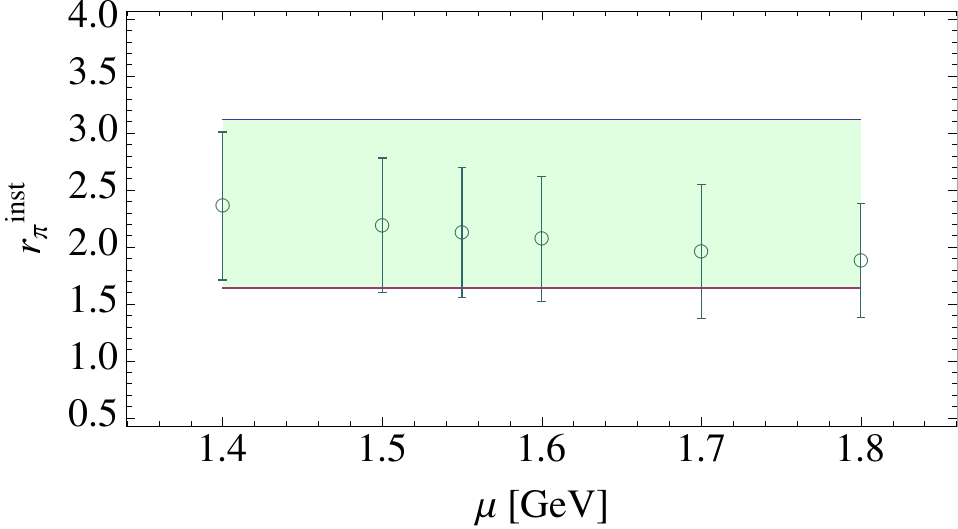}}
\caption{
\scriptsize 
 {\bf a)} $\tau$-behaviour  of  $r_\pi$ for  $\mu=1.55$ GeV and for different values of $t_c$ from the instanton sum rule. 
 {\bf b)} $\mu$-behaviour  of  the optimal value of $r_\pi$ deduced from a) and sources of errors analogue to the ones in Fig. \ref{fig:rpi}.
}
\label{fig:rpinst}
\end{center}
\end{figure} 
\nin
  \subsection*{\b Convergence of the  QCD series}
  \nin
 Here, we study the convergence of the PT series in the more general case where $\mu$ is  arbitrary and not necessarily correlated to the value of $\tau$. The particular case $\mu=\tau^{-1/2}$ often used in the literature will be also discussed in the next paragraph.
 
-- We study in Fig \ref{fig:conv}a, the convergence of the PT QCD series at the value of the subtraction scale $\mu =1.55$ GeV  where a $\mu$ stability is obtained (inflexion point in Fig. \ref{fig:rpi}b) for the  ratio ${\cal R}^{\pi}_5 (\tau,\mu)$ used for the estimate of $r_\pi$ (lower family of curves). One can notice that for $\tau\approx 0.6$ GeV$^{-2}$ where a $\tau$ inflexion is obtained (Fig. \ref{fig:rpi}a), the $\alpha_s, \alpha^2_s, \alpha^3_s$ and $\alpha^4_s$ effects are respectively $-7.5, -9.4, -9.0$, and $-4.5\%$ of the preceding PT series: LO, up to NLO, up to N2LO up to N3LO and up to N4LO or equivalently, the PT series behaves as: $1-0.07-0.08-0.08-0.05$. The convergence of the PT series is slow but each corrections to $r_\pi$ are reasonably small. However, one can notice that the convergence of the PT series is much better here than in the case (often used in the literature)  $\mu=\tau^{-1/2}$ where the $\tau$ stability is obtained at larger value of $\tau=1.9$ GeV$^{-2}$ (Fig. \ref{fig:rpimutau}a) as we shall discuss later on. 

-- We show in Fig \ref{fig:conv}b, the convergence of the power corrections for $r_\pi$ (lower family of curves). We see that the $d=2,~4, ~6$ dimension operator effects are $-1.4, -4.2$ and $-1.4\%$ of the preceding sum of contributions or equivalently, the NP series normalized to the PT series behaves as: $1-0.018-0.065-0.046$ indicating a slow convergence of the OPE but relatively small corrections. 
\begin{figure}[hbt] 
\begin{center}
\centerline 
{\includegraphics[width=8cm  ]{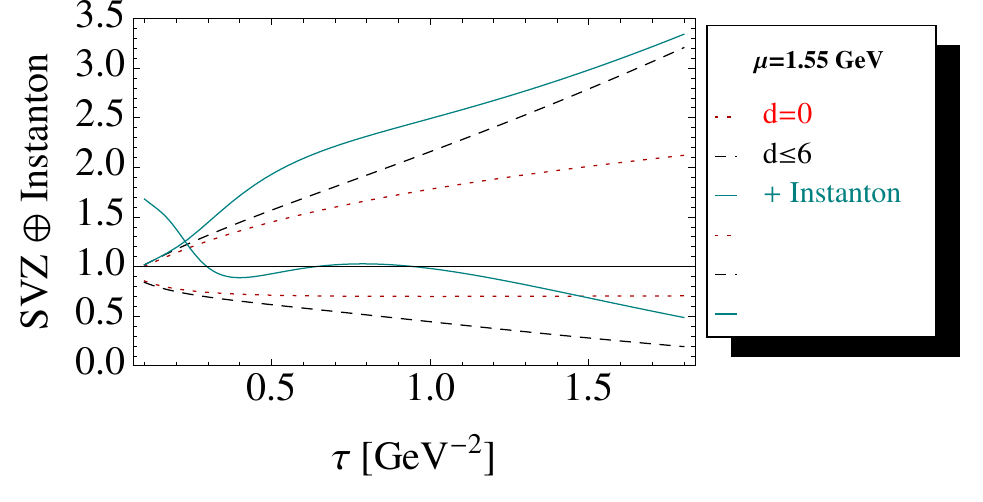}}
\caption{
\scriptsize 
$\tau$-behaviour  of  the NP corrections including the instanton contribution to $\sqrt{\cal L}_5^{\pi}(\tau,\mu)$ (upper group of curves) and of ${\cal R}^{\pi}_5 (\tau,\mu)$ (lower group of curves)  appropriately normalised to 1 for $\tau=0$ and using $\mu=1.55$ GeV. 
}
\label{fig:convinst}
\end{center}
\end{figure} 
\nin
  \subsection*{\b Tachyonic gluon mass and large order PT-terms to  $r_\pi$ }
  \nin
The tachyonic gluon mass decreases the value of $r_\pi$ by about 0.1 which is relatively negligible.  The smallness of this contribution is consistent with the small contribution of the estimated N5LO terms using a geometric growth of the PT series. Using the duality between the long PT series and the short PT series $\oplus$ $1/Q^2$ correction \cite{SNZ}, the inclusion of the $1/Q^2$ into the OPE mimics the contributions of the non-calculated higher order in the PT series which are
therefore expected to be small numbers. Here and in the following the  PT series is truncated at N4LO (order $\alpha_s^4$) and the sum of $\alpha_s^n$ corrections for $n\geq 5$ is approximated by the tachyonic gluon mass contribution. 
  \subsection*{\b $r_\pi$ from instanton sum rule at arbitrary $\mu$}
  \nin
 We include  the instanton contribution into the OPE using the expression given in Eq.~(\ref{eq:instanton}). The variations of $r_\pi$ versus $\tau$  and $t_c$ for different values of $\mu$  are similar to the one in Fig. \ref{fig:rpi}a. The optimal result is obtained for $\tau\simeq (0.9\pm 0.1)$ GeV$^{-2}$ (inflexion point) and $t_c\simeq 2.25$ GeV$^2$ (minimum in $t_c$). Comparing the behaviour of the curves using the SVZ and the SVZ $\oplus$ instanton expansion in Fig. \ref{fig:rpinst}a, one can notice that the instanton contribution has shifted the inflexion point at slightly higher $\tau$-values. Normalized to the PT contributions, the sum of the SVZ term to ${\cal R}^{\pi}_5 (\tau,\mu)$ at $\tau\simeq 0.9$ GeV$^{-2}$ is -33\% of the PT contribution while the one of the instanton is about +77\% (see Fig. \ref{fig:convinst}).  At $\mu=1.55$ GeV and for $\tau=(0.9\pm 0.1)$ GeV$^{-2}$, we obtain after an iteration procedure by using the final value of $\la \bar dd\ra$ condensate obtained in Eq. (\ref{eq:dd}) for the $d=4$ condensate contribution\,\footnote{The value of  the four-quark condensate extracted from the V and V+A channels  quoted in Table \ref{tab:param} is expected \cite{NASON,SNTAU9} to be weakly  affected by instanton in the OPE.}:
 \bea
r_\pi^{inst}&=&2.11(1)_{\Lambda}(3)_{\lambda^2}(10)_{\bar uu}(10)_{G^2}(1)_{\bar uGu}(17)_\rho(1)_{\rho_c}\nnb\\
&&(50)_{\Gamma_{\pi}}(3)_{t_c}(5)_\tau\nnb\\
&=&2.11\pm 0.57~.
\label{eq:rpinsta}
\eea
 We study in Fig. \ref{fig:rpinst}b the $\mu$ behaviour of the optimal results in the range $\mu=1.4$ to 1.8 GeV like in the previous analysis. 
Then, we deduce the mean value:
\beq
r_\pi^{inst}=2.06\pm 0.55\pm 0.20_{syst}~,
\label{eq:rpinst}
\eeq
where the first error comes from the most precise determination, while the systematic error  is  the distance of the mean from it.

  \subsection*{\b  $r_\pi$ from LSR at $\mu=\tau^{-1/2}$}
\begin{figure}[hbt] 
\begin{center}
\centerline {\hspace*{-8.5cm} a) }\vspace{-0.6cm}
{\includegraphics[width=8.5cm  ]{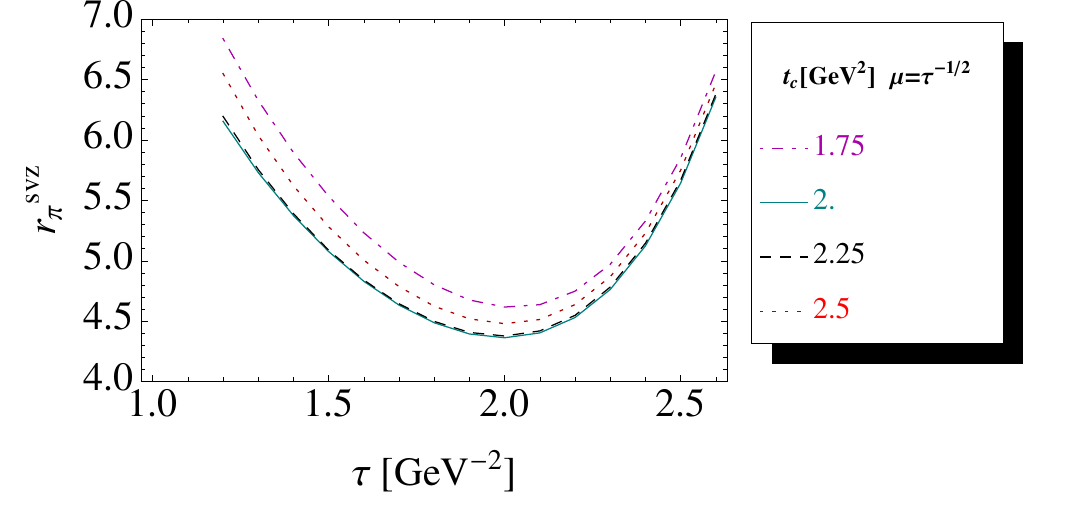}}
\centerline {\hspace*{-8.5cm} b) }\vspace{-0.3cm}
{\includegraphics[width=8.5cm  ]{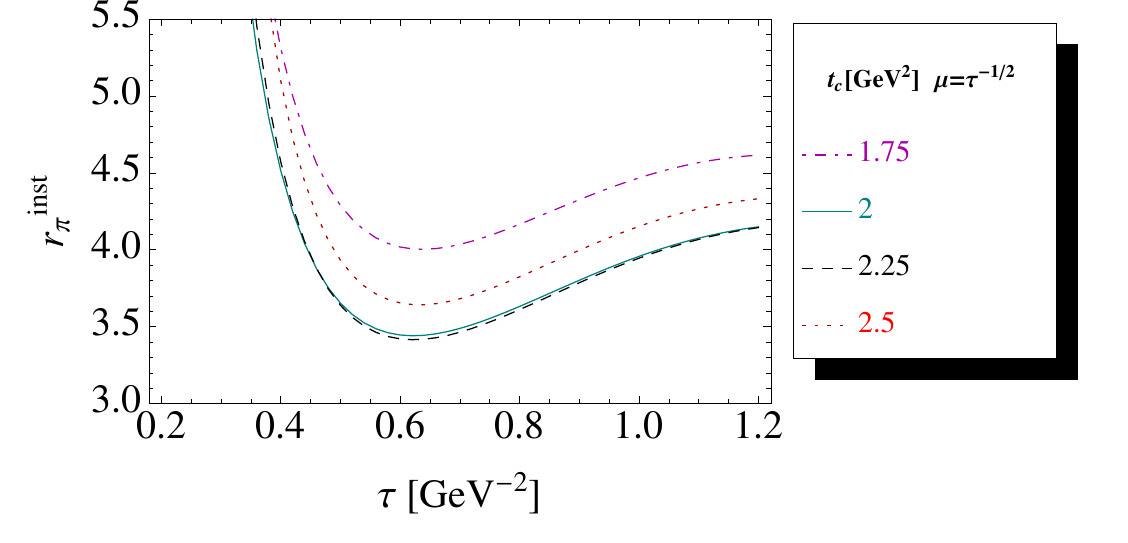}}
\caption{
\scriptsize 
{\bf a)} $\tau$-behaviour  of  $r_\pi$ for the case $\mu=\tau^{-1/2}$ in the
case of the SVZ expansion; {\bf b)} the same as in a) but for  SVZ $\oplus$ instanton contribution.
}
\label{fig:rpimutau}
\end{center}
\end{figure}

  \nin
We complete the analysis in the case where the subtraction constant $\mu$ 
is equal to the sum rule variable $1/\sqrt{\tau}$. This case is interesting as it
does not possess the Log$^n\mu^2\tau$ terms appearing in the PT series which
have large coefficients and which are now absorbed into the running of $\alpha_s(\tau)$
from the renormalization group equation. This case has been largely used in the literature
(for reviews see, e.g.: \cite{SNB0,SNB1,SNB2,SNB3}). The analysis is very similar to the
previous case. In Fig.  \ref{fig:rpimutau}, we show the $\tau$-behaviour of the results  in the
case of the SVZ expansion and SVZ $\oplus$ instanton contribution where in both cases a minimum in $t_c$ is obtained.  
\nin
We obtain for $\tau=(2\pm 0.1)$ GeV$^{-2}$ and $t_c=2$ GeV$^2$. 
\bea
r_\pi^{svz}&=&4.36(120)_{\Lambda}(12)_{\lambda^2}(81)_{\bar uu}(67)_{G^2}(1)_{\bar uGu}(169)_\rho\nnb\\
&&(56)_{\Gamma_{\pi}}(2)_{t_c}(4)_\tau\nnb\\
&=&4.36\pm2.39~,
\label{eq:rpitaumu_svz}
\eea
For the instanton case, we obtain for $\tau=(0.6\pm 0.1)$ GeV$^{-2}$ and $t_c=2.25$ GeV$^2$. :
\bea
r_\pi^{inst}&=&3.40(8)_{\Lambda}(1)_{\lambda^2}(7)_{\bar uu}(14)_{G^2}(1)_{\bar uGu}(15)_\rho(22)_{\rho_c}\nnb\\
&&(104)_{\Gamma_{\pi}}(2)_{t_c}(4)_\tau\nnb\\
&=&3.40\pm 1.09~.
\label{eq:rpitaumu_inst}
\eea
  \subsection*{\b Final result and comparison with some existing predictions}
\begin{figure}[hbt] 
\begin{center}
\centerline 
{\includegraphics[width=9cm  ]{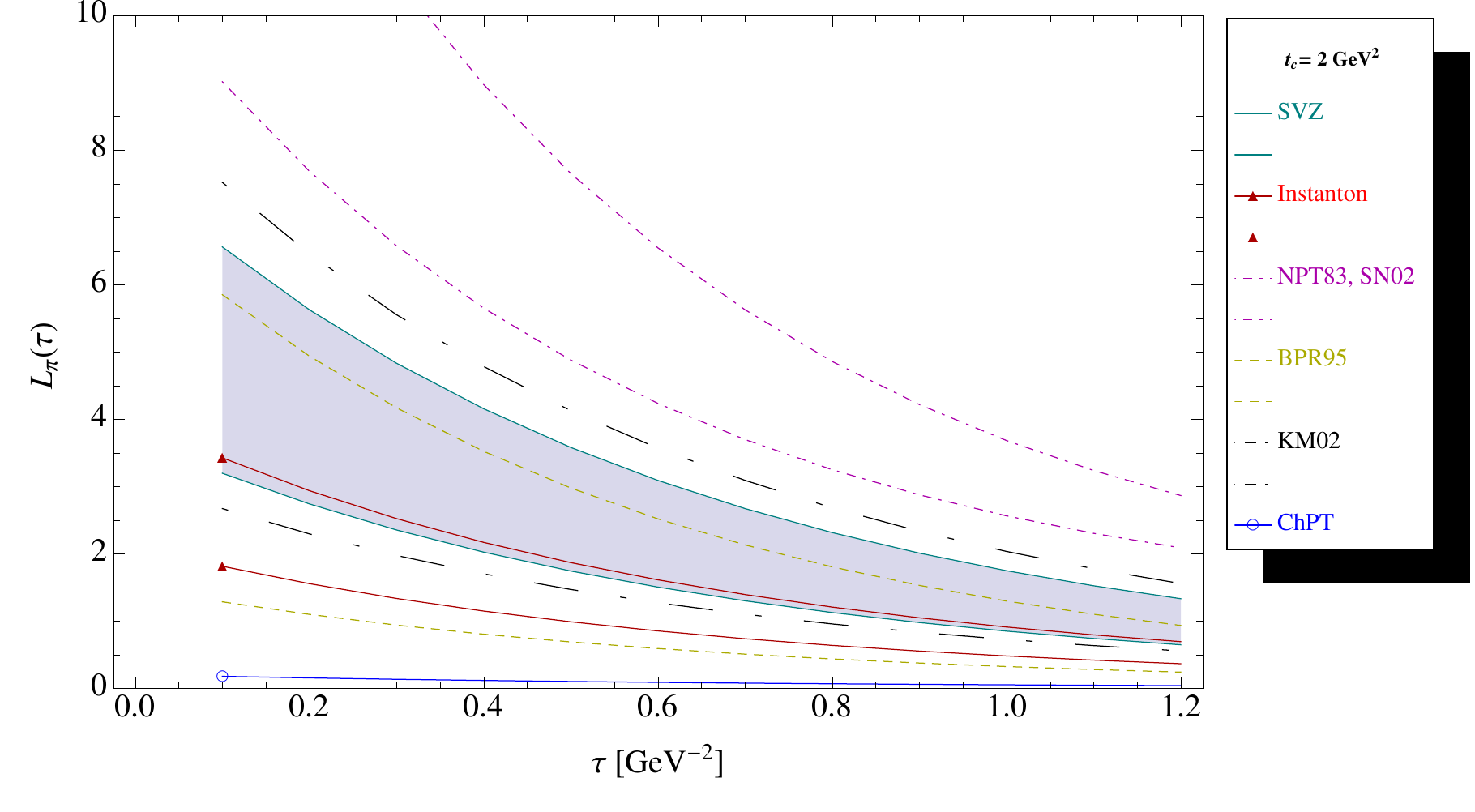}}
\caption{
\scriptsize 
Comparison of some other determinations of  $r_\pi$ for a given value of $t_c=5$ GeV$^2$ which corresponds to the optimal value of $(m_u+m_d)$. The blue continuous line with a circle is the ChPT prediction without a resonance.The results of NPT83 \cite{TREL} and SN02 \cite{SNB1} are within a narrow width approximation. The errors due to the experimental width of the $\pi(1300)$ have been introduced in the result of BPR95 \cite{BPR}. }
\label{fig:pionexp}
\end{center}
\end{figure}

  \nin
 One can remark a nice agreement within the errors between the different results in Eq. (\ref{eq:rpi_ope}) with Eq. (\ref{eq:rpitaumu_svz}) and Eq. (\ref{eq:rpinst}) with Eq. (\ref{eq:rpitaumu_inst}). However, the estimates from the LSR with $\mu=\tau^{-1/2}$ are obtained at much larger values of $\tau$ and are sensitive to the NP contributions rendering the estimate less accurate.  
Taking the mean of the previous estimates, we deduce our final results:
\bea
 r_\pi^{svz}&=&4.42\pm 1.56~\lrar  {f_{\pi'}\over f_\pi}=\ga 2.42\pm 0.43\dr 10^{-2},\nnb\\
 r_\pi^{inst} &=&2.36\pm 0.74~\lrar{ f_{\pi'}\over f_\pi}=\ga 1.77\pm 0.28\dr 10^{-2},
\label{eq:rpifinal}
\eea
where we have separated the determinations from the SVZ and SVZ $\oplus$ instanton 
sum rules. The errors come from the most precise estimate to which we have added a systematics from the  distance of the mean to it. 
In Fig. \ref{fig:pionexp}, we compare the above two results,  with the existing ones in the current literature: NPT83 \cite{TREL}, SN02 \cite{SNB1}, BPR95 \cite{BPR}, KM02 \cite{MALT} for the quantity:
\beq
L_\pi(\tau)\equiv r_\pi BWI_0~,
\label{eq:piprim}
\eeq
involved in the Laplace sum rule estimate of $(m_u+m_d)$  which we shall discuss in the next sections. 
 Here $BWI_0$ defined in Eq. (\ref{eq:bwi}) is the integrated spectral function entering into the lowest moment Laplace sum rule ${\cal L}_5^\pi(\tau)$.  For this comparison, we have used:
 
-- $r_\pi=(9.5\pm 2.5)$ and consistently the NWA for the results in NPT83 and SN02 from \cite{SNB1} (see also \cite{LARIN2}). 

-- For KM02 \cite{MALT}, we add coherently the $\pi(1300)$ and $\pi(1800)$ contributions which may be an overestimate as they may have a destructive interference like in BPR95 \cite{BPR}. We use the decay constants $f_\pi(1300)=(2.2\pm 0.57)$ MeV and $f_\pi(1800)=(1.36\pm 0.21)$ MeV obtained in \cite{MALT}  and consistently a Breit-Wigner parametrization of the spectral function. 

-- For BPR95, we add, into their parametrization, the error due to the $\pi(1300)$ width which is not included in their original work and we use the mixing parameter $\zeta=0.0.234+i~0.1$ between the $\pi(1300)$ and $\pi(1800)$ which reproduces the best fit to the experimental curves which observe the $\pi(1.8)$ in hadronic interactions. 

-- The CHPT parametrization from BPR95 \cite{BPR} without any resonance is also given in  Fig. \ref{fig:pionexp}.

The results obtained in \cite{BPR} and \cite{MALT} are model-dependent as they depend on the way of treating the $\pi(1800)$ contribution into the spectral function. One can see explicitly in Fig. \ref{fig:rpi}c that the $\pi(1800)$ contribution to $r_\pi$ is negligible rendering the result in this paper less model-dependent thanks to the exponential weight of the LSR which safely suppresses its contribution. 

From the previous comparison, we notice that the prediction from the SVZ expansion has a better agreement (within the errors) with the predictions shown in Fig. \ref{fig:pionexp}, while the one including the instanton tends to underestimate the $\pi(1300)$ contribution to the spectral function. One can also notice that the NWA used in NPT83 \cite{TREL} and SN02 \cite{SNB1} used there tends to give larger values which presumably indicates that the NWA  is not sufficient for a good description of the pseudoscalar spectral function. 

\nin
\section{ Estimate of $( m_u+m_d)$ within the SVZ expansion}
  \subsection*{\b The LSR  for arbitrary  $\mu$}
\begin{figure}[hbt] 
\begin{center}
\centerline {\hspace*{-8.5cm} a) }\vspace{-0.6cm}
{\includegraphics[width=8cm  ]{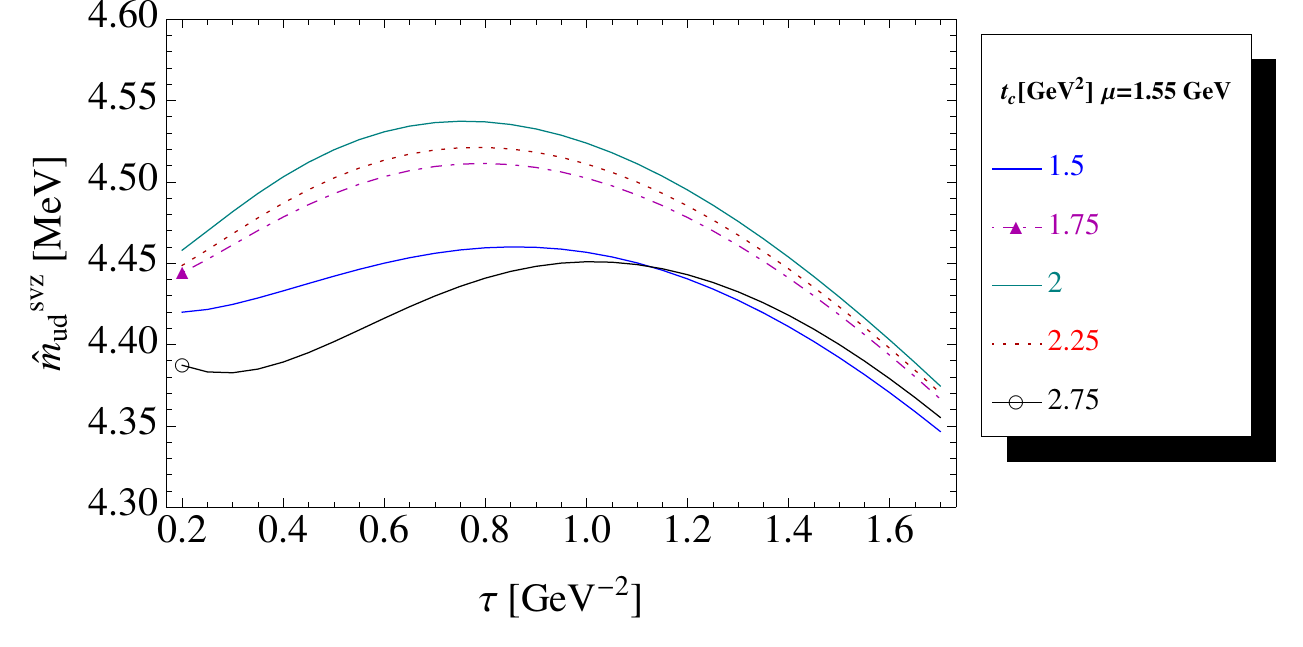}}
\centerline {\hspace*{-8.5cm} b) }\vspace{-0.3cm}
{\includegraphics[width=7cm  ]{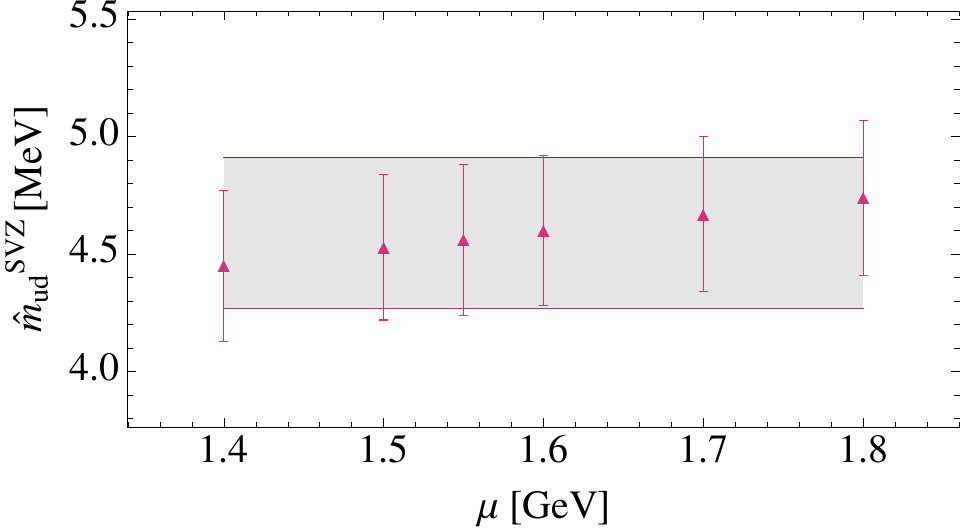}}
\caption{
\scriptsize 
$\tau$- and $t_c$-dependence of $\hat m_{ud}$ from the Laplace sum in Eq.~(\ref{eq:lsr}) at the subtraction scale $\mu=1.55$ GeV. The filled coloured region corresponds to mean value
where the errors come from the most precise determination $ \oplus$ systematics. }
\label{fig:mudtau}
\end{center}
\end{figure}

  \nin
  We find convenient to extract the RGI scale independent mass  defined in Eq. (\ref{eq:rgi}):
 \beq
 \hat m_{ud}\equiv {1\over 2} (\hat m_u+\hat m_d)
 \eeq
 from the Laplace sum rule ${\cal L}_5^\pi(\mu,\tau)$ in Eq.~(\ref{eq:lsr}). The QCD expression of ${\cal L}_5^\pi(\mu,\tau)$ is given in Eq. (\ref{eq:lsrqcd}). We shall use into the spectral function, parametrized as in Eq. (\ref{eq:duality}), the value of $r_\pi$ obtained in Eq. (\ref{eq:rpifinal}) and
we do not transfer the QCD continuum contribution to the QCD side of the sum rule. In this way, we obtain a much better $\tau$-stability but we have  an initial value of $ \hat m_{ud}$ for quantifying the QCD continuum contribution. Therefore, we use an obvious iteration procedure by replacing successively the initial input value of $\hat m_{ud}$ with the obtained value and so on. The procedure converges rapidly after 2 iterations. 
 We show, in Fig. \ref{fig:mudtau}a, the $\tau$-dependence of $ \hat m_{ud}$ for different values of $t_c$ and for a given value of the subtraction point $\mu=1.55$ GeV, where the optimal value of $r_\pi$ has been obtained. One can find from this figure that one obtains a  $\tau$-stability for $\tau\simeq (0.7\pm 0.1)$ GeV$^{-2}$. A minimum in $t_c$ is also obtained for $t_c \simeq (2\sim 2.25)$ GeV$^2$ consistent with the one in the determination of $r_\pi$ . Using the values of the parameters in Table \ref{tab:param}, we extract the optimal value of the sum of the RGI $u,d$ quark masses for $\mu$=1.55 GeV and at the extrema (stability region) of the curve:
 \bea
\hat m_{ud}^{svz}&=&4.56(11)_{\Lambda}(6)_{\lambda^2}(1)_{\bar uu}(10)_{G^2}(0)_{\bar uGu}(7)_\rho\nnb\\
&&(10)_{\Gamma_{\pi}}(27)_{r_\pi}(0)_{\pi(1.8)}(0)_\tau(2)_{t_c}~{\rm MeV}~,\nnb\\
&=&(4.56\pm 0.32)~{\rm MeV}~,
\label{eq:mud1}
\eea
 where the errors due to the localisation of the $\tau$ and $t_c$ stability region are negligible like also the $\pi(1.8)$ contribution using its coupling from \cite{MALT}. 
\nin
We study the $\mu$-dependence of the result in Fig. \ref{fig:mudtau}b and deduce the mean value:
\beq
\la\hat m_{ud}^{svz}\ra=(4.59\pm 0.31\pm 0.06_{syst})~{\rm MeV}~,
\label{eq:mud}
\eeq
which corresponds to the one of the slight $\mu$ inflexion point obtained  around $(1.55-1.60)$ GeV. 
The first error is the one from the most precise measurement.
The second one is a systematics coming from the distance of the mean to its  central value.
  \subsection*{\b $\hat m_{ud}$ from the Laplace sum rule at $\mu=\tau^{-1/2}$}
\begin{figure}[hbt] 
\begin{center}
{\includegraphics[width=8cm  ]{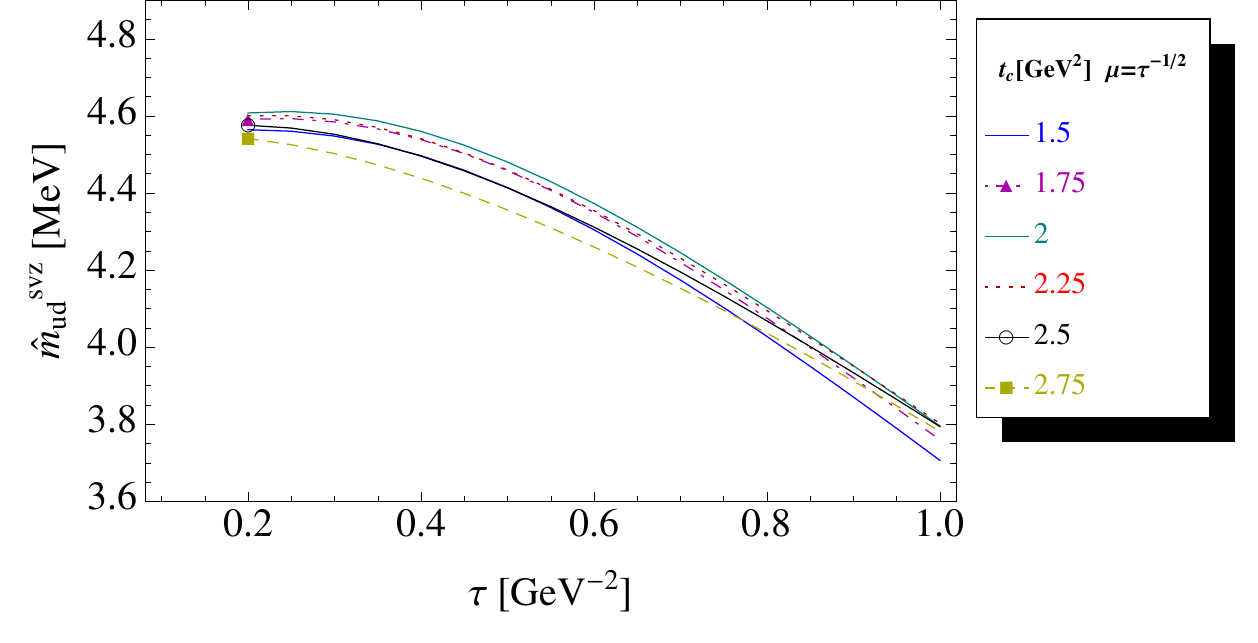}}
\caption{
\scriptsize 
$\tau$- and $t_c$-dependence of $\hat m_{ud}$ from the Laplace sum in Eq.~(\ref{eq:lsr}) at the subtraction scale $\mu=\tau^{-1/2}$.
}
\label{fig:mudtaumu}
\end{center}
\end{figure}

  \nin
As mentioned previously, this sum rule has been widely used in the current literature for extracting $m_{ud}$. We shall use it here as another method for determining $m_{ud}$. The analysis is similar to the one for arbitrary $\mu$. We show the $\tau$-dependence for different $t_c$ in Fig. \ref{fig:mudtaumu}. One can see that unlike the case of arbitrary $\mu$, there is no $\tau$-stability here. 
Therefore, we shall not consider  this approach in this analysis. 
  \subsection*{\b Convergence of the QCD series}
  \nin
 Like in the case of $r_\pi$, we study the convergence of the PT series and of the OPE. 
 
-- We shall study the different contributions of the truncated PT series to $\sqrt{\cal L}_5^\pi(\mu,\tau)$ for $\mu= 1.55$ GeV where a  $\mu$ stability is obtained (slight inflexion point in Fig. \ref{fig:rpi}b). The relative strengths of each truncated conributions are given in Fig \ref{fig:conv}a (upper family of curves). One can deduce that for $\tau\approx  0.7$  GeV$^{-2}$ where the $\tau$-stability is obtained (Fig. \ref{fig:mudtau}), the $\alpha_s, \alpha^2_s, \alpha^3_s$ and $\alpha^4_s$ effects are respectively $+29, +13, +6$ and $+3\%$ of the preceding PT series: LO, NLO, N2LO and N3LO, which is equivalent to: $1+0.29+0.17+0.09+0.05$ when normalized to the LO PT contribution. It indicates a good convergence of the PT series.  

-- We show in Fig \ref{fig:conv}b, the convergence of the power corrections for $\sqrt{\cal L}_5^\pi(\mu,\tau)$ for $\mu=1.55$ GeV (upper family of curves). We see that the $d=2,4,6$ contributions are $+3.8, +4.8$ and $+2.9\%$ of the preceding sum of contributions ($PT, ~PT~ \oplus ~d=2, PT~ \oplus~ d=2+4$) or equivalently: $1+0.04+0.05+0.03$ when normalized to the PT series indicating a slow convergence but relatively small corrections.  

\nin
  \subsection*{\b Tachyonic gluon mass and large order PT-terms to  $\hat m_{ud}$}
  \nin
If we do not include the tachyonic gluon mass contribution into the SVZ expansion, the value of $\hat m_{ud}$
obtained in Eq. (\ref{eq:mud1})  would increase by 0.15 MeV which is relatively negligible confirming again the good convergence of the PT series if one evokes a duality between the tachyonic gluon mass and the not yet calculated higher order PT corrections \cite{SNZ}. Within this duality argument, one can estimate  the contribution of the large order non calculated PT terms
(sum of the higher order $\alpha_s^n:~n\geq 5$)  by the one of the tachyonic gluon mass. 
  \subsection*{\b Final estimate of $\hat m_{ud}$ within the SVZ expansion }
  \nin
We consider, as a {\it  final estimate} of $\hat m_{ud}$ within the SVZ expansion, the mean value   in  Eq. (\ref{eq:mud}) which is:
\beq
\la \hat m_{ud}^{svz}\ra=(4.59\pm 0.32)~{\rm MeV}.
\label{eq:mudsvz}
\eeq
\section{  $m_{ud}$ from the  instanton Laplace sum rules}
\begin{figure}[hbt] 
\begin{center}
\centerline {\hspace*{-8.5cm} a) }\vspace{-0.6cm}
{\includegraphics[width=8.5cm  ]{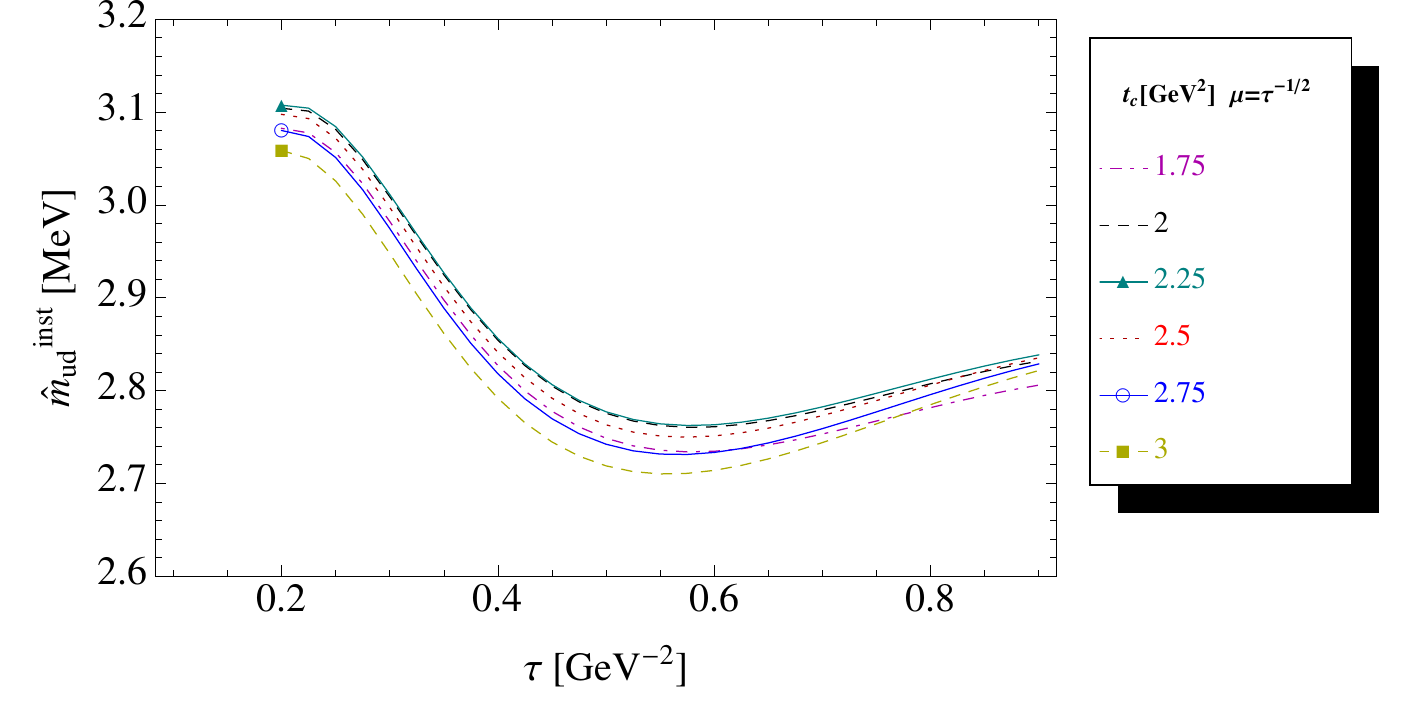}}
\centerline {\hspace*{-8.5cm} b) }\vspace{-0.3cm}
{\includegraphics[width=8.5cm  ]{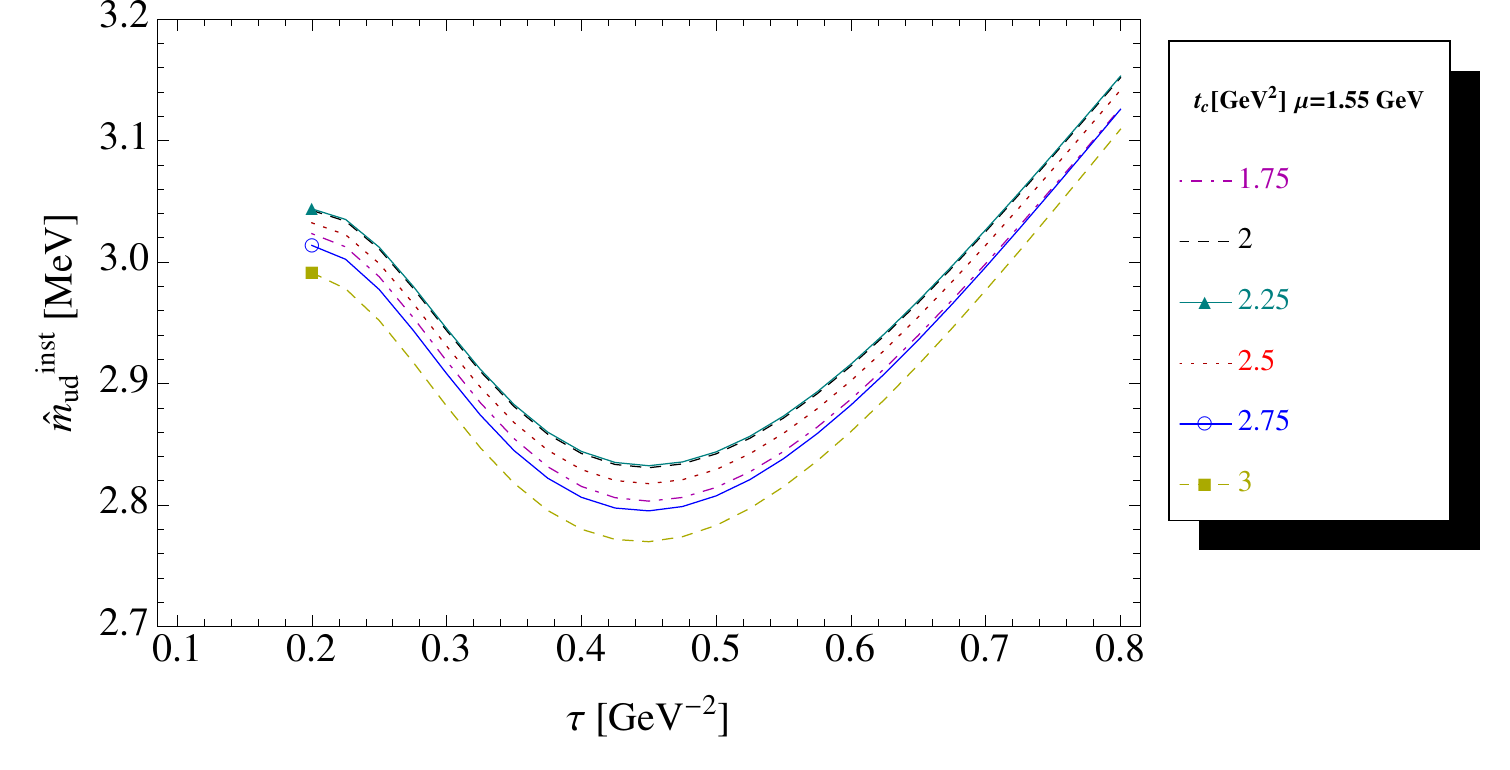}}
\centerline {\hspace*{-8.5cm} c) }\vspace{-0.3cm}
{\includegraphics[width=7.5cm  ]{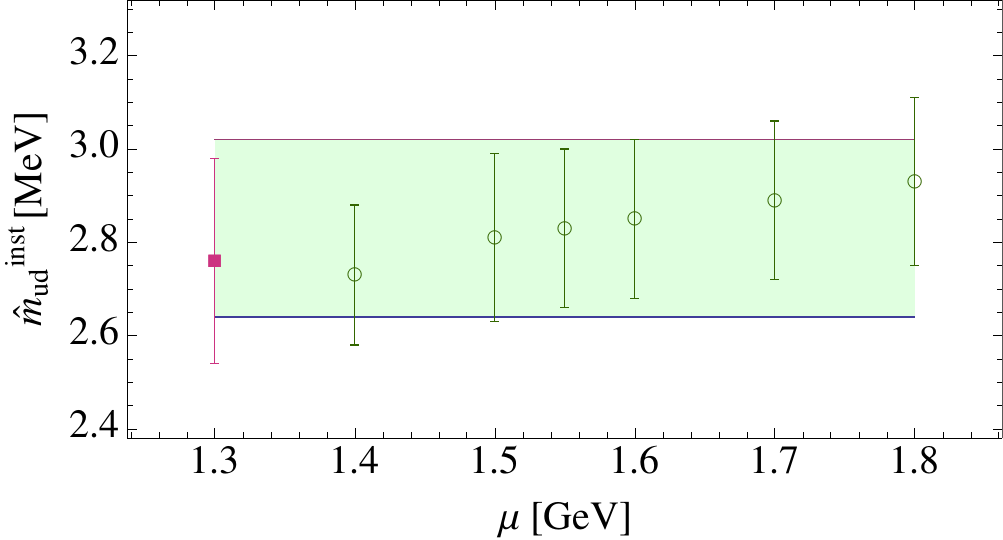}}
\caption{
\scriptsize 
{\bf a)} $\tau$-behaviour  of  $m_{ud}$ in the case $\mu=\tau^{-1/2}$;  {\bf b)} the same as in a) but in the case $\mu=1.55$ GeV;  {\bf c)} $\mu$-behaviour of $\hat m_{ud}$ obtained from LSR: the red box is the value from $\mu=\tau^{-1/2}$ in b). Same meaning of coloured region as in Fig. \ref{fig:mudtau}b.
}
\label{fig:mdinst}
\end{center}
\end{figure}

For optimizing the instanton contribution, we work at the same subtraction point $\mu=(1.4-1.8)$ GeV where $r_\pi^{inst}$ has been obtained. We shall use the value of $r_\pi$ extracted in Eq. (\ref{eq:rpifinal}). 
 We repeat the previous analysis by taking into account the instanton contribution. 
Its contribution to $\sqrt{\cal L}_5^\pi(\mu,\tau)$ compared to the OPE up to d=6 condensates is shown in Fig. \ref{fig:convinst} (upper family of curves) for $\mu=1.55$ GeV where the estimate of $r_\pi$ has been also optimized. For $\tau\approx (0.4\sim 0.5)$ GeV$^{-2}$where the sum rule is optimized (Fig. \ref{fig:mdinst}a and Fig. \ref{fig:mdinst}b for $\mu=\tau^{-1/2}$ and $\mu=1.55$ GeV), the instanton contribution is  about +36\% (resp 15\%) of the perturbative $\oplus$  $d\leq 6$ condensates for the sum rule subtracted at $\mu=\tau^{-1/2}$ (resp. $\mu=1.55$ GeV).   $t_c$-stability is  reached for $t_c\simeq (2\sim 2.25)$ GeV$^2$. We deduce respectively from the sum rule subtracted at $\mu=\tau^{-1/2}$ and $\mu=1.55$ GeV:
\bea
\hat m_{ud}^{inst}\vert_{1.55}
&=&2.81(4)_{\Lambda}(2)_{\lambda^2}(0)_{\bar uu}(1)_{G^2}(0)_{\bar uGu}(1)_\rho\nnb\\
&&(7)_{\rho_c}(6)_{\Gamma_{\pi}}(13)_{r_\pi}(1)_\tau(0)_{t_c}~{\rm MeV}~,\nnb\\
&=&(2.81\pm 0.17)~{\rm MeV}~,\nnb\\
\hat m_{ud}^{inst}\vert_{\tau^{-1/2}}&=&2.76(6)_{\Lambda}(2)_{\lambda^2}(0)_{\bar uu}(2)_{G^2}(0)_{\bar uGu}(1)_\rho\nnb\\
&&(6)_{\rho_c}(6)_{\Gamma_{\pi}}(19)_{r_\pi}(2)_\tau(0)_{t_c}~{\rm MeV}~,\nnb\\
&=&(2.76\pm 0.22)~{\rm MeV}~.
\eea
We show in Fig. \ref{fig:mdinst}c the $\mu$ behaviour of the different determinations from which we deduce the mean  value with the conservative error:
\beq
 \la \hat m_{ud}^{inst}\ra\simeq (2.81\pm 0.16\pm 0.10_{syst})~{\rm MeV}~,
 \label{eq:mdinst}
 \eeq
 which we consider as a determination of $\hat m_{ud}$ from the SVZ $\oplus$ instanton sum rules. 
\nin
  \section{$\hat m_{ud}$ and $\overline{m}_{ud}$(2)  from Laplace sum rules}
  \nin
We consider, as a {\it  final estimate} of the RGI mass $\hat m_{ud}$, the  results  obtained in  Eqs. (\ref{eq:mudsvz})  
 and (\ref{eq:mdinst}) from which  we deduce the running masses at order $\alpha_s^4$ evaluated at 2 GeV in units of MeV:
\beq
\overline{m}_{ud}^{svz}=3.95\pm 0.28~,~~~~~~~~\overline{m}_{ud}^{inst}=2.42\pm 0.16~.
\label{eq:mud_final}
\eeq
We have not taken the mean value of the two results taking into account the controversial contribution of the instanton into the pseudoscalar sum rule \cite{SHURYAK,IOFFE,NASON,ZAKI}.
The value of the sum $\overline{m}_{ud}$ within the SVZ expansion agrees  within the errors with the average: 
$
\overline{m}_{ud}=(5.0\pm 0.9)~{\rm MeV}~,
$
quoted in \cite{SNB1} and co\"\i ncide with the one $
\overline{m}_{ud}=(3.95\pm 0.30)~{\rm MeV}~,
$
deduced by combining the value of the average $\overline m_s$ 
from different phenomenological sources  and the ChPT mass ratio $m_s/m_{ud}$ \cite{SNmass}.
We consider the previous results as improvements of our previous determinations in \cite{SNB1} from the Laplace pseudoscalar sum rules and some other sum rules determinations in this channel compiled in PDG13 \cite{PDG}. The previous results can also be compared with the recent lattice average \cite{LATT13}: 
\beq
\overline{m}_{ud}^{latt}=(3.6\pm 0.2)~{\rm MeV}~~ [{\rm resp.}~~ (3.4\pm 0.1)~{\rm MeV}]~,
\eeq
obtained using $n_f=2$ [resp. $n_f=2+1$] dynamical fermions where there is a good agreement within the errors
with the SVZ value but not with the instanton one which is lower. 
Using the mean of the range of different results quoted in PDG13 \cite{PDG} for the ratio:
\beq
{m_u\over m_d}=0.50(3)~,
\label{eq:ratiomass}
\eeq
which (a priori) does not favour the solution $m_u=0$ advocated in connection with the strong CP-problem (see e.g \cite{MANOHAR}), 
one can deduce the value of the $u$ and $d$ running quark masses at 2 GeV in units of MeV:
\bea
\overline{m}_{u}^{svz}&=&2.64\pm 0.28~,~~~~~~~~~\overline{m}_{u}^{inst}=1.61\pm 0.14\nnb\\
\overline{m}_{d}^{svz}&=&5.27\pm 0.49~,~~~~~~~~~ \overline{m}_{d}^{inst}=3.23\pm 0.29~.
\label{eq:mfinal}
\eea
Using the GMOR relation in Eq. (\ref{eq:gmor}), we can deduce the value of the running light quark condensate: $\la\bar uu\ra\simeq \la\bar dd\ra$ at 2 GeV in units of MeV$^3$:
\beq
-\la\overline{\bar dd}\ra^{svz}=(276\pm 7)^3,~~~~~~~~~ -\la\overline{\bar dd}\ra^{inst}=(325\pm 7)^3,
\label{eq:dd}
\eeq
and to the spontaneous mass in units of MeV defined in Eq.~(\ref{eq:rgi}):
\beq
\mu_d^{svz}=253\pm 6~, ~~~~~~~~~~~~~~~~~\mu_d^{inst}=298\pm 7~,
\label{eq:dbd}
\eeq
where the SVZ result is in perfect agreement with the one from \cite{SNmass} used 
in Table \ref{tab:param}. 
The results are summarized in Table \ref{tab:res}. 
\section{Laplace sum rule estimate of $f_{K'}$}
\begin{figure}[hbt] 
\begin{center}
\centerline {\hspace*{-8.5cm} a) }\vspace{-0.6cm}
{\includegraphics[width=8cm  ]{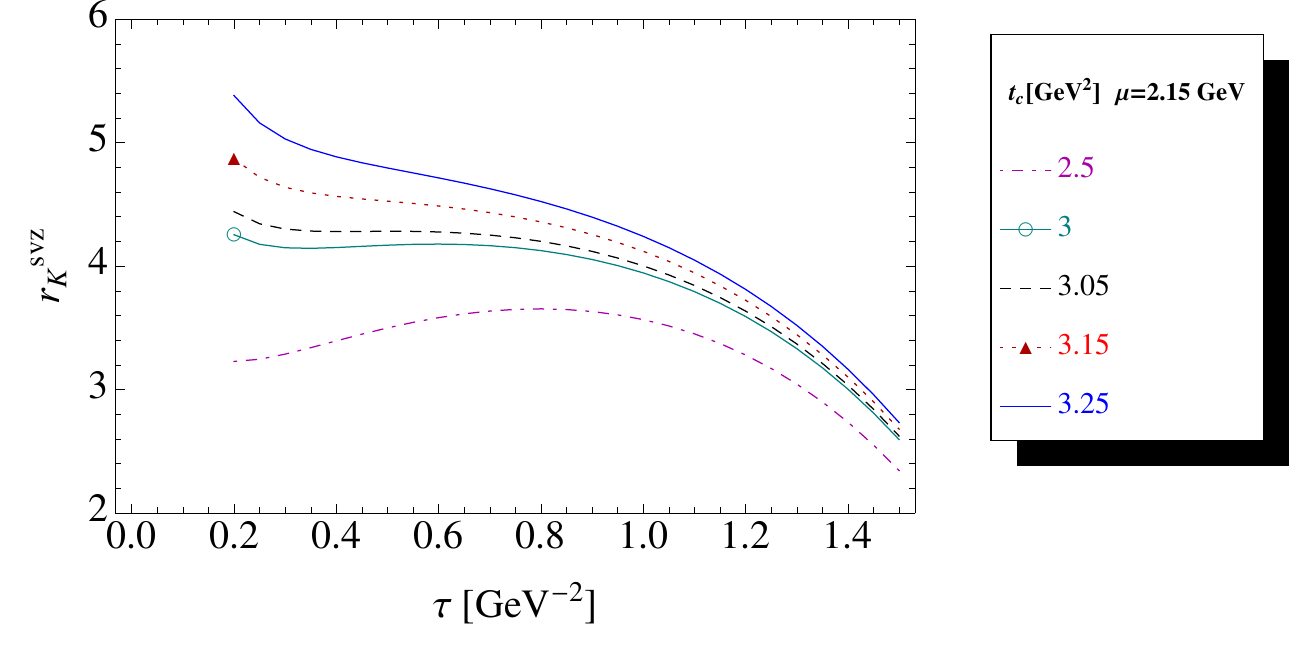}}
\centerline {\hspace*{-8.5cm} b) }\vspace{-0.3cm}
{\includegraphics[width=7cm  ]{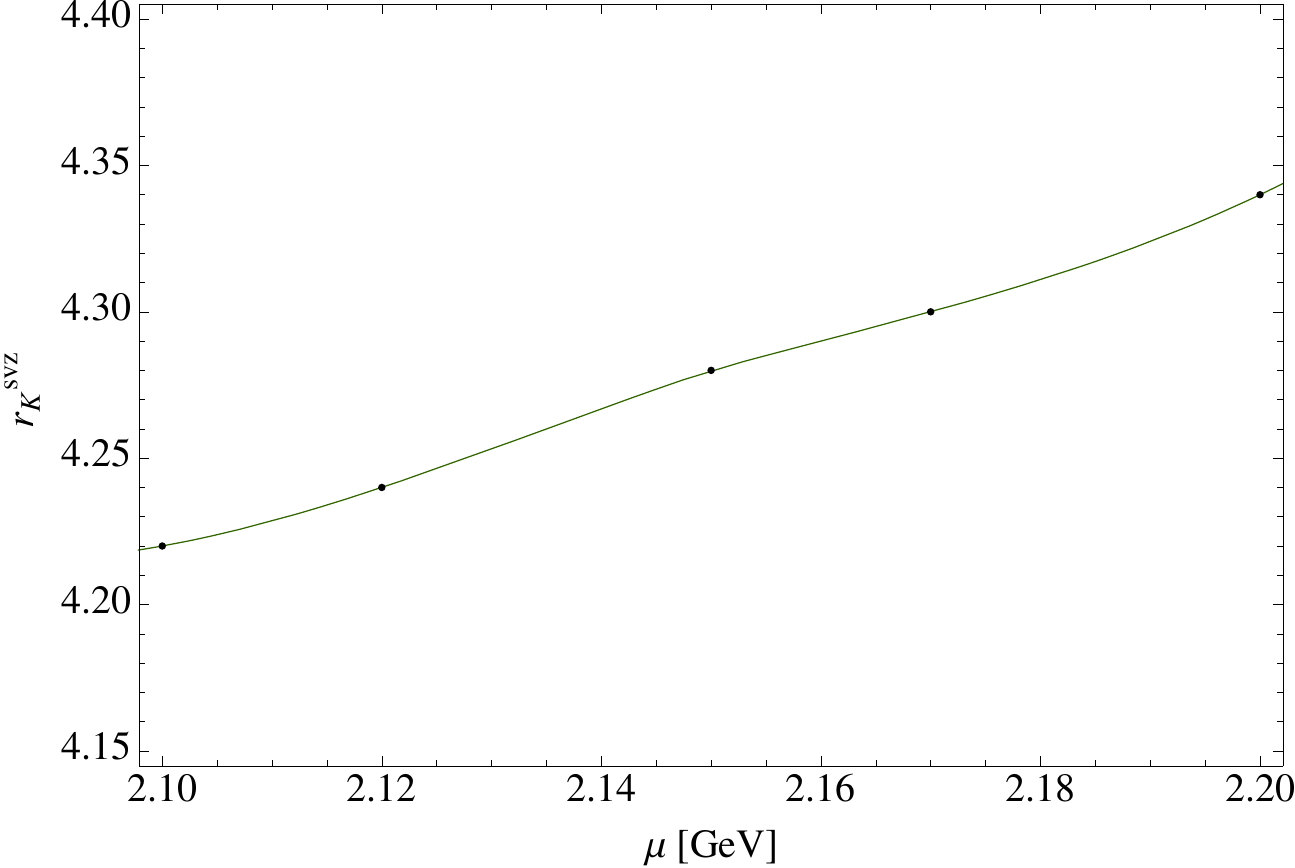}}
\centerline {\hspace*{-8.5cm} c) }\vspace{-0.3cm}
{\includegraphics[width=7cm  ]{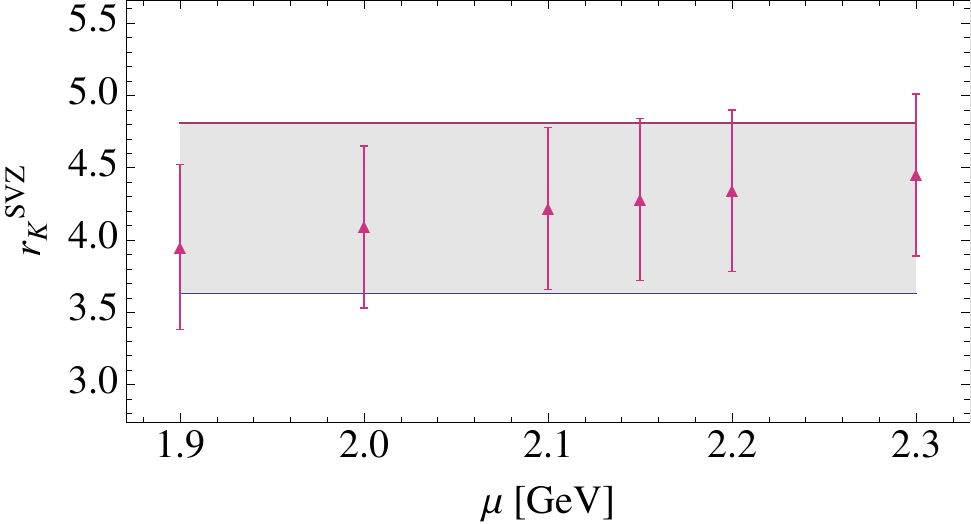}}
\caption{
\scriptsize 
{\bf a)} $\tau$-behaviour  of  $r_K$ for a given value $\mu=2.15$ GeV of subtraction point and for different values of $t_c$. 
; {\bf b)} $\mu$-behaviour  of  the optimal central values of $r_K$ deduced from a)
; {\bf c)} $\mu$-behaviour  and mean value of $r_K$ from LSR: same meaning of coloured region  as in Fig. \ref{fig:rpi}b.
}
\label{fig:rk}
\end{center}
\end{figure} 
\nin
Using the same method as in the case of the $\pi'$, we shall estimate the $K'\equiv K(1460)$ decay constant through:
\beq
r_K\equiv {M^4_{K'}f_{K'}^2\over m^4_Kf_K^2}~.
\eeq 
  \subsection*{\b Analysis within the SVZ expansion for arbitrary $\mu$}
  \nin
 
-- We show the $\tau$- and $t_c$-behaviours of $r_K$ in Fig. \ref{fig:rk}a for a given value of the subtraction point $\mu=2.15$ GeV, where the a $\tau$-maximum is obtained at $ 0.7$ GeV$^{-2}$ and an almost plateau from $\tau \simeq (0.6\sim 0.9)$ GeV$^{-2}$ for $t_c\simeq (3.05\pm 0.10)$ GeV$^2$.  At this scale, one can inspect using curves similar to  Fig. \ref{fig:conv}a  (lower families of curves) that the PT corrections to ${\cal R}^{K}_5 (\tau,\mu)$ are small: the $\alpha_s, \alpha^2_s, \alpha^3_s$ and $\alpha^4_s$ effects are respectively $-6., -6.4, -6.8$ and $-6.1\%$ of the preceding PT series including: LO, NLO, N2LO and N3LO contributions which is equivalent to: $1-0.06-0.06-0.06-0.05$ when normalized to the LO perturbative series. The PT series converges slowly but the corrections are small. The NP corrections  similar to the ones in Fig. \ref{fig:conv}b remain reasonably small: the $d=2,4, 6$ dimension operators contributions are $-1.3, -10.5$ and -4.3\% of the preceding sum of contributions ($PT, ~PT~ \oplus ~d=2, PT~ \oplus~ d=2+4$) which is equivalent to: $1-0.01-0.10-0.04$ when normalized to the PT contributions. 

-- In Fig. \ref{fig:rk}b, we show the $\mu$-behaviour of the central values of the optimal results obtained from Fig. \ref{fig:rk}a.  One can notice a slight inflexion point like in the case of the pion. At this point $\mu=2.15$ GeV, we obtain:
\bea
r_K^{svz}&=&4.28(6)_{\Lambda}(6)_{\lambda^2}(11)_{\bar uu}(31)_{G^2}(1)_{\bar uGu}(25)_\rho\nnb\\
&&(1)_{m_s}(^{-24}_{+11})_\kappa(30)_{\Gamma_{K}}(10)_{t_c}\nnb\\
&=&4.28\pm 0.56~.
\label{eq:rk_ope1}
\eea
In Fig. \ref{fig:rk}c, we study the effects of $\mu$ by moving it from 1.9 to 2.3 GeV around the inflexion point. 
The average of these results leads to the final estimate:
\beq
r_K^{svz}=4.22\pm 0.54\pm 0.23_{syst}\lrar {f_{K'}\over f_K}=(23.5\pm 1.6)10^{-2},
\label{eq:rk_ope}
\eeq
where one can remark from Eq. \ref{eq:rpifinal} that $r_\pi\approx r_K$ as expected from chiral symmetry arguments.
  \subsection*{\b Analysis within the SVZ expansion for $\mu=\tau^{-1/2}$}
\begin{figure}[hbt] 
\begin{center}
\centerline 
{\includegraphics[width=8cm  ]{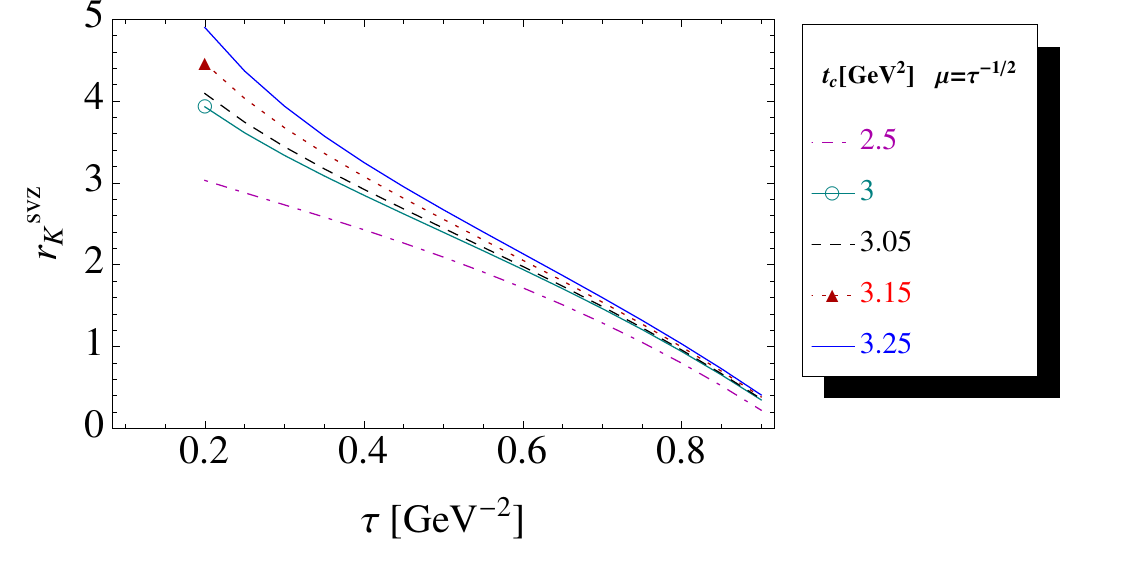}}
\caption{
\scriptsize 
 $\tau$-behaviour  of  $r_K$ for $\mu=\tau^{-1/2}$  and for different values of $t_c$.}
\label{fig:rkmutau}
\end{center}
\end{figure}

  \nin
We show the result of the analysis in Fig. \ref{fig:rkmutau} where there is no $\tau$ stability.
Therefore, we shall not consider the result of this sum rule in the following. 
\nin

  \subsection*{\b $r_K$ from instanton sum rules}
  \nin
 The analysis of the sum rule  for arbitrary $\mu$ does not lead to a conclusive result.
The one for $\mu=\tau^{-1/2}$ is given in Fig. \ref{fig:rkinst}. Like in the case of the pion, we shall use the value of the $\la \bar dd\ra$ condensate obtained in Eq. (\ref{eq:dd}) for the $d=4$ condensate contribution and the value in Table \ref{tab:param} for the four-quark condensate extracted from the V and V+A channels which is weakly affected by instanton effects \cite{NASON,SNTAU9}. We deduce for $t_c=(3.0-3.05)$ GeV$^2$:
\begin{figure}[hbt] 
\begin{center}
{\includegraphics[width=8cm  ]{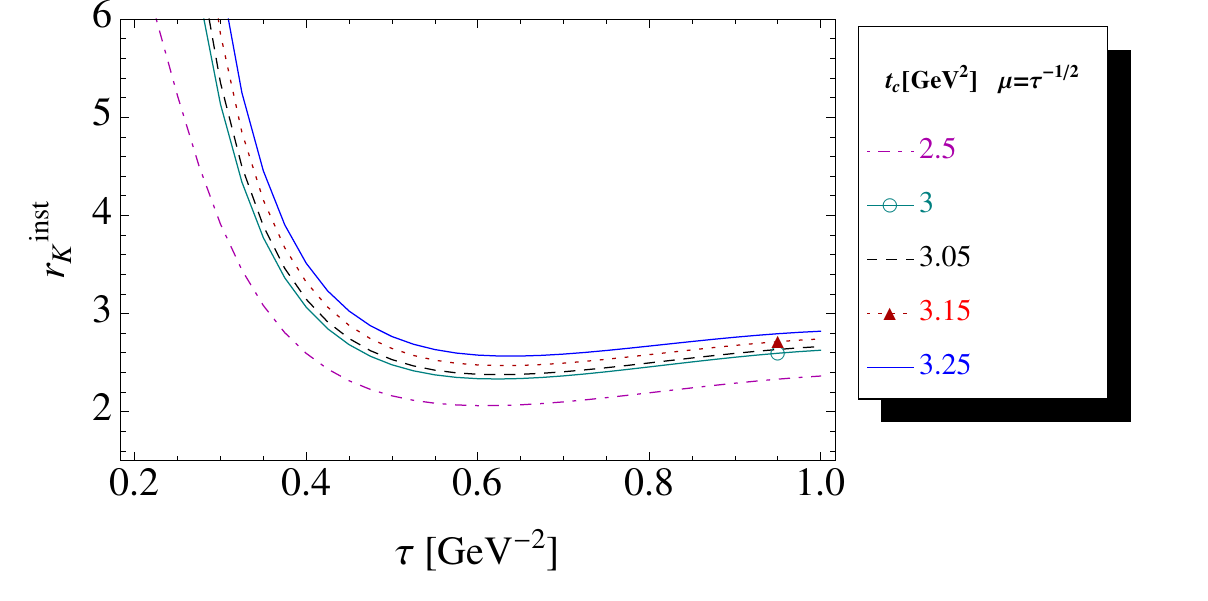}}
\caption{
\scriptsize 
$\tau$-behaviour  of  $r_K$ for a given value $\mu=\tau^{-1/2}$ of the subtraction point and for different values of $t_c$ from the instanton sum rule}
\label{fig:rkinst}
\end{center}
\end{figure} 
\nin
\bea
r_K^{inst}&=& 2.36(17)_{svz}(16)_{\rho_c}(12)_{\Gamma_K}(2)_{t_c}(0)_\tau\nnb\\
&=&2.36\pm 0.26\lrar {f_{K'}\over f_K}=(17.6\pm 1.0)10^{-2}.
\label{eq:rk_inst}
\eea
where the index SVZ means that the corresponding error is the quadratic sum of the ones due to the PT contributions and to the NP terms within the SVZ expansion defined in Section 2:
\beq
(17)_{svz}\equiv (6)_{\Lambda}(0)_{\lambda^2}(4)_{\bar uu}(10)_{G^2}(2)_{\bar uGu}(9)_\rho
(1)_{m_s}(6)_\kappa~.
\eeq
  \subsection*{\b Comparison with some other predictions}
\begin{figure}[hbt] 
\begin{center}
\centerline 
{\includegraphics[width=8cm  ]{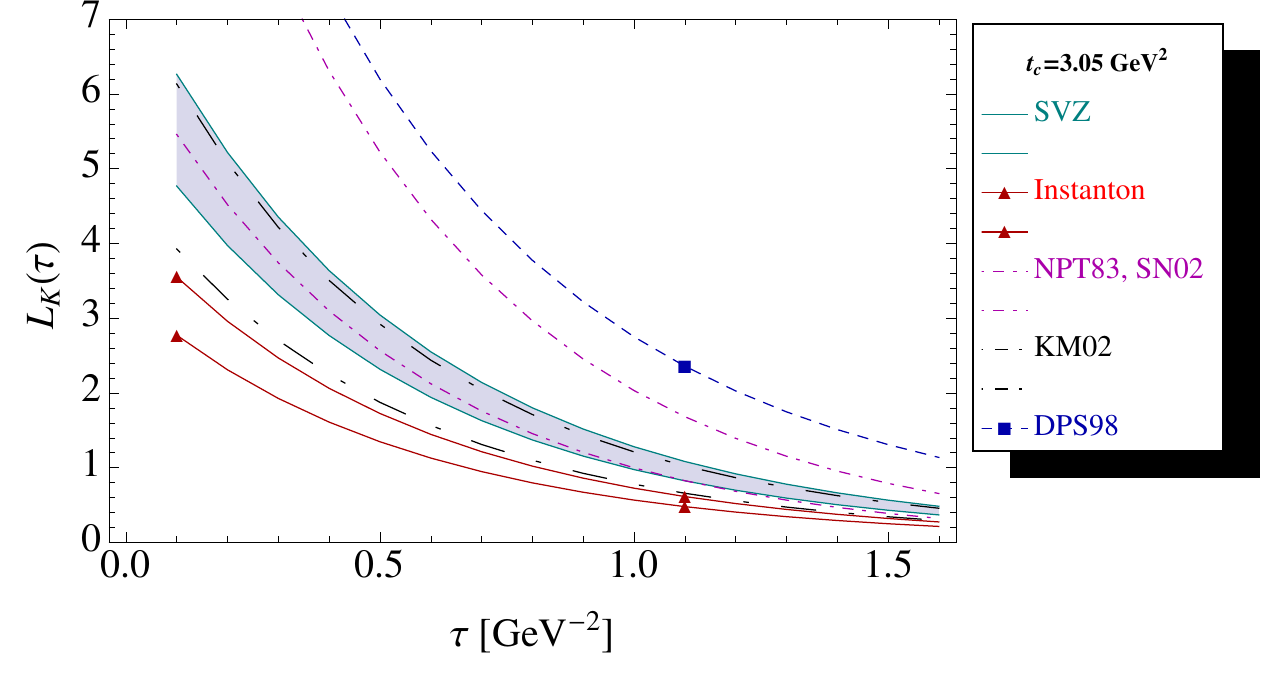}}
\caption{
\scriptsize 
Comparison of our determination of  $r_K$ from SVZ and instanton sum rules
with the ones in the current literature: NPT83 \cite{TREL}, SN02 \cite{SNB1}, KM02 \cite{MALT} and DPS98 \cite{DOM}). We use  $t_c=3.05$ GeV$^2$. }
\label{fig:kaonexp}
\end{center}
\end{figure}

  \nin
We compare in Fig. \ref{fig:kaonexp},  our results from Eqs. (\ref{eq:rk_ope}) and (\ref{eq:rk_inst}) for:
\beq
L_K(\tau)\equiv r_K BWI_0~,
\label{eq:kprim}
\eeq
with the existing ones in the current literature (NPT83 \cite{TREL}, SN02 \cite{SNB1}, KM02 \cite{MALT} and DPS98 \cite{DOM}). The results of NPT83 \cite{TREL} and SN02 \cite{SNB1} are obtained within a narrow width approximation. The ones of KM02 \cite{MALT} and DPS98 \cite{DOM} include finite width correction. There are fair agreement between different determinations with the exception of the one from \cite{DOM} which is relatively high (central value shown in Fig. \ref{fig:kaonexp}). This high value may be either due to the coherent sum and equal coupling of the $K(1460)$ and $K(1800)$ contribution assumed in the amplitude or due to an overall normalization factor\,\footnote{Notice that instead of \cite{DOM} in the kaon channel, a destructive interference has been assumed by \cite{BPR} in the pion channel, with which agrees our estimate in the pion channel.}. We also see in Fig. \ref{fig:kaonexp} that the instanton sum rule estimate is relatively small compared with the one from the sum rule within the SVZ expansion and with some other determinations. As we shall see later, this low value of the $K(1460)$ contribution will imply a smaller value of $m_s$ from the instanton sum rule estimate. 
\nin
\section{ Laplace sum rule estimate of $\hat{m}_{us}$ and  $\overline{m}_{us}$}
Defining:
\beq
m_{us}=( m_u+m_s)~,
\eeq
we now turn to the estimate of the RGI $ \hat m_{us}$ and running $ \overline{m}_{us}$
 sum of masses. 

  \subsection*{\b   $\hat m_{us}$ within the SVZ expansion for arbitrary $\mu$}
  \nin
\begin{figure}[hbt] 
\begin{center}
\centerline {\hspace*{-8.5cm} a) }\vspace{-0.6cm}
{\includegraphics[width=8cm  ]{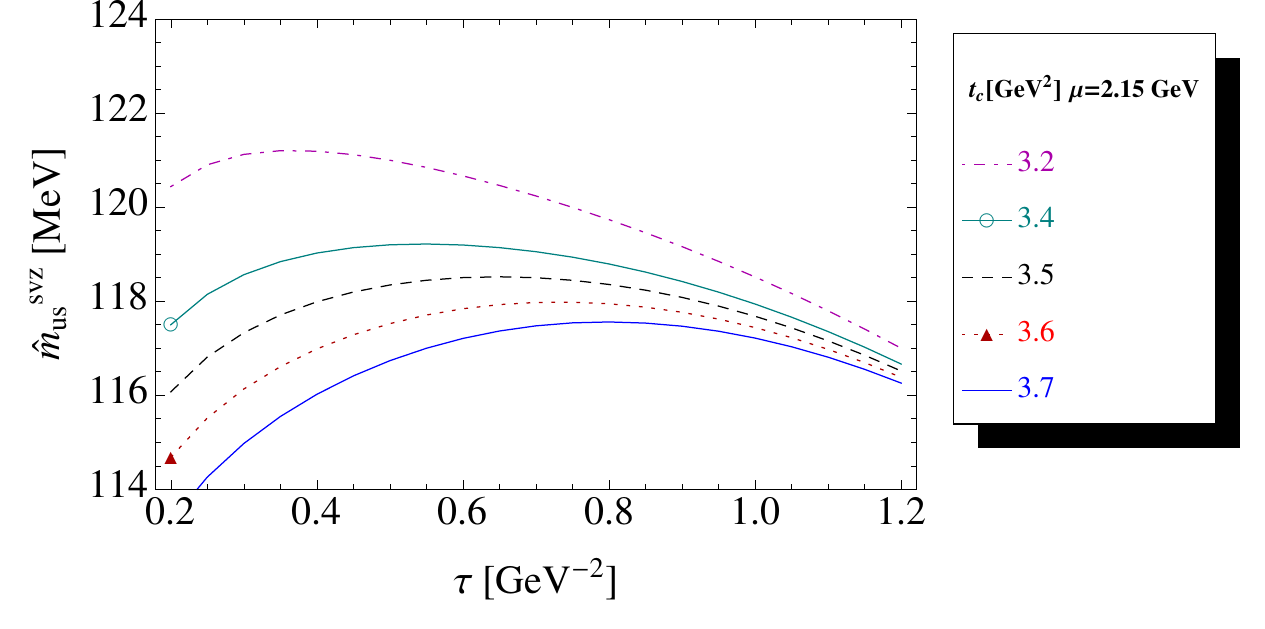}}
\centerline {\hspace*{-8.5cm} b) }\vspace{-0.3cm}
{\includegraphics[width=7.5cm  ]{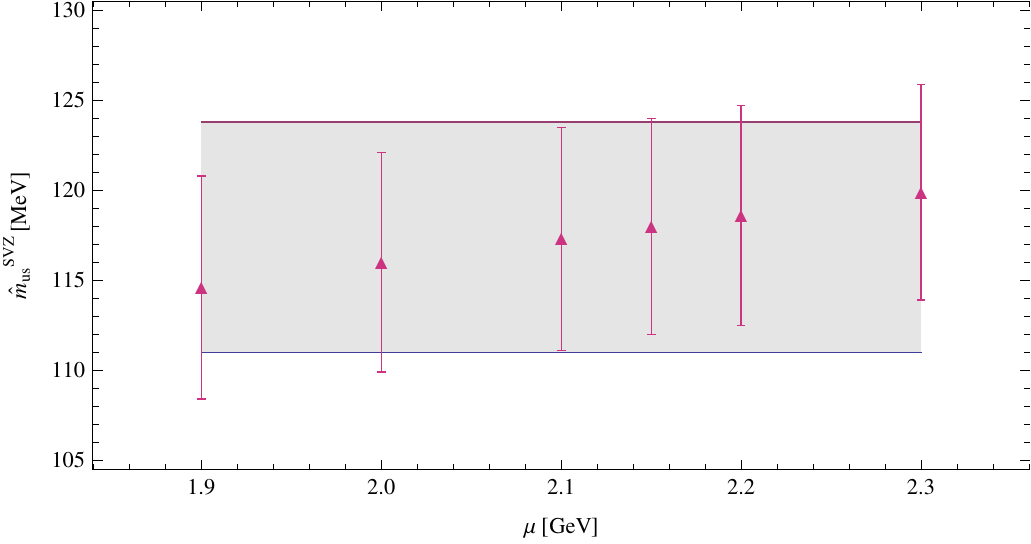}}
\caption{
\scriptsize 
{\bf a)} $\tau$-behaviour  of  $\hat m_{us}$ for a given value $\mu=2.15$ GeV of subtraction point and for different values of $t_c$. 
; {\bf b)} $\mu$-behaviour  of  the optimal value of  $\hat m_{us}$ deduced from a): same meaning of coloured region as in Fig. \ref{fig:mudtau}b. }
\label{fig:ms}
\end{center}
\end{figure} 
\nin
 We show in Fig. \ref{fig:ms}a the $\tau$-behaviour  of  $\hat m_{us}$ for a given value $\mu=2.15$ GeV of the subtraction point and for different values of $t_c$ where we have used  the value of $r_K$ in Eq. (\ref{eq:rk_ope}). The largest range of $\tau$-stability of about $(0.6-0.9)$ GeV$^{-2}$ is reached at $t_c\simeq (3.6\pm 0.1)$ GeV$^2$. However, one can notice that the value of $t_c$  corresponding to the best stability for $\hat m_{us}$ differs slightly with the one $t_c\simeq (3.05\pm 0.10)$ GeV$^2$ for $r_K$. We consider this systematics by enlarging the range of $t_c$ to $t_c\simeq (3.6\pm 0.4)$ GeV$^2$. Using the initial value of $m_{us}$ in Table 1 and after two obvious iterations, we obtain for $\mu=2.15$ GeV:
 \bea
  \hat m^{svz}_{us}&=&118.0(30)_{\Lambda}(16)_{\lambda^2}(6)_{\bar uu}(23)_{G^2}(1)_{\bar uGu}(13)_\rho\nnb\\
&&(5)_\kappa (6)_{\Gamma_{K}}(40)_{r_K}(4)_{t_c}(0)_\tau~{\rm MeV}\nnb\\
&=&(118.0\pm 6.0)~{\rm MeV}~,
\label{eq:ms_ope1}
\eea
where the main error comes from the $K'$-meson contribution. 
 We show in Fig. \ref{fig:ms}b the $\mu$-behaviour  of  the optimal value of  $\hat m_{us}$  from Fig. \ref{fig:ms}a from which we deduce the mean value:
 \beq
 \la  \hat m^{svz}_{us}\ra=(117.4\pm 5.9\pm 2.5_{syst})~{\rm MeV}~.
 \label{eq:ms_ope}
 \eeq
  \subsection*{\b   $\hat m_{us}$ within the SVZ expansion for  $\mu=\tau^{-1/2}$}
  \nin
We redo the previous analysis but for $\mu=\tau^{-1/2}$. Unfortunately, we have no stability like in the case of $r_K$.
  \subsection*{\b $\hat m_{us}$ from the instanton sum rule at arbitrary $\mu$}
\begin{figure}[hbt] 
\begin{center}
\centerline {\hspace*{-8.5cm} a) }\vspace{-0.6cm}
{\includegraphics[width=8cm  ]{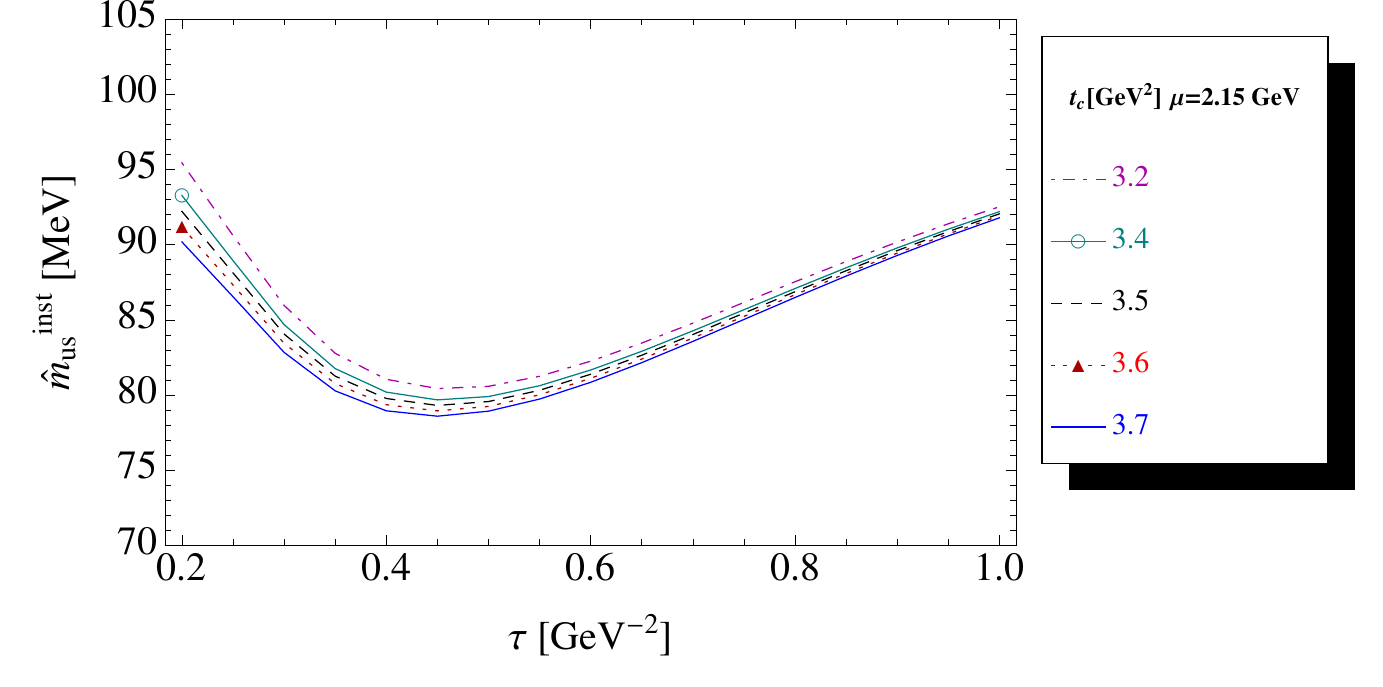}}
\centerline {\hspace*{-8.5cm} b) }\vspace{-0.3cm}
{\includegraphics[width=6cm  ]{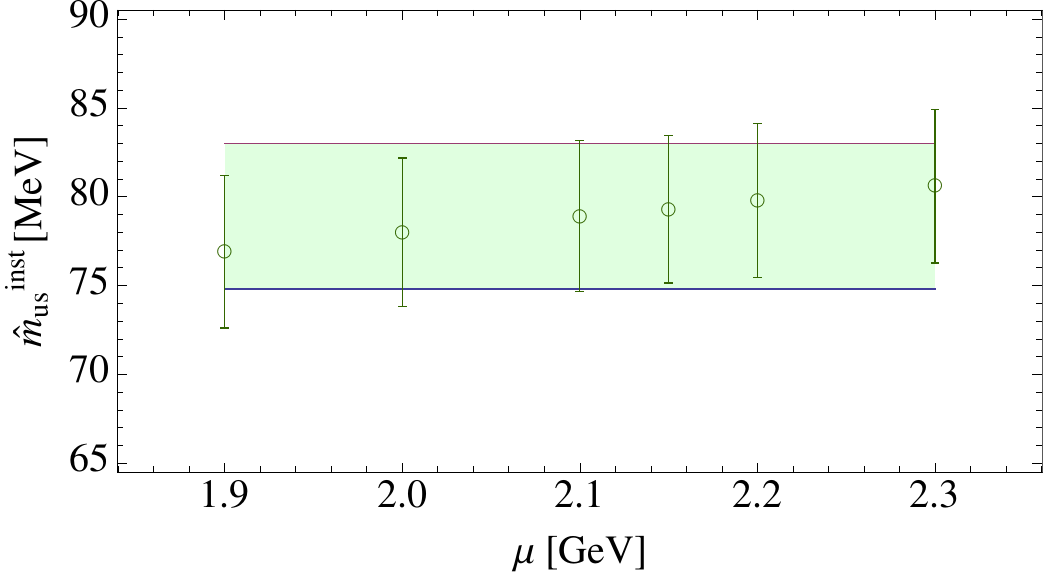}}
\caption{
\scriptsize 
{\bf a)} $\tau$-behaviour  of  $\hat m_{us}$ from the instanton sum rule for a given value $\mu=2.15$ GeV of the subtraction point and for different values of $t_c$; {\bf b)}  $\mu$-behaviour of the optimal results: same meaning of coloured region as in Fig. \ref{fig:mudtau}b.}
\label{fig:msinst}
\end{center}
\end{figure}

  \nin
We show in Fig. \ref{fig:msinst} the $\tau$-behaviour  of  $\hat m_{us}$ from the instanton sum rule at different values of $t_c$ and for a given value $\mu=2.15$ GeV. We take the optimal value at the $\tau$-minimum of about $(0.45\pm 0.10)$ GeV$^{-2}$ where we notice that the effect of  $t_c$ in the range  $(3.6\pm 0.4)$ GeV$^2$
is relatively small. We obtain:
\bea
\hat m_{us}^{inst}&=&79.3(16)_{\Lambda}(7)_{\lambda^2}(2)_{\bar uu} (5)_{G^2}(0)_{\bar uGu} (1)_\rho\nnb\\
&&(24)_{\rho_c}(2)_\kappa(5)_{\Gamma_{K}}(20)_{r_K}(15){t_c}(12)_\tau
~{\rm MeV}\nnb\\
&=&(79.3\pm 4.1)~{\rm MeV}~.
\label{eq:msinst1}
\eea
We study the $\mu$-dependence of the results in Fig. \ref{fig:msinst} from which we deduce the mean value:
 \beq
  \hat m^{inst}_{us}=(78.9\pm 4.1\pm 0.4_{syst})~{\rm MeV}~.
\label{eq:msinst}
 \eeq
\nin
  \subsection*{\b $\hat m_{us}$ from the instanton sum rule at $\mu=\tau^{-1/2}$}
\begin{figure}[hbt] 
\begin{center}
{\includegraphics[width=8cm  ]{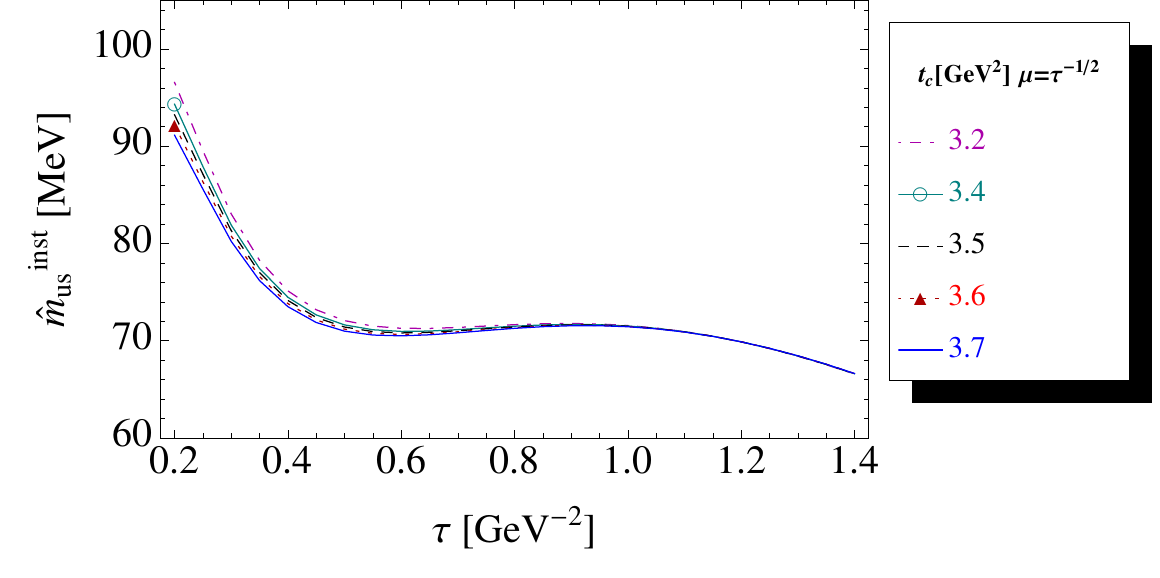}}
\caption{
\scriptsize 
$\tau$-behaviour  of  $\hat m_{us}$ from the instanton sum rule  for $\mu=\tau^{-1/2}$ and for different values of $t_c$.  }
\label{fig:mstaumu_inst}
\end{center}
\end{figure}

  \nin
 The analysis is shown in Fig. \ref{fig:mstaumu_inst}. Our optimal results correspond to $\tau=(0.5\sim 0.9)$ GeV$^{-2}$ and $t_c=(3.6\pm 0.4)$ GeV$^2$. We deduce
  \bea
  \hat m^{inst}_{us}&=&70.65(132)_{\Lambda}(65)_{\lambda^2}(14)_{\bar uu} (70)_{G^2}(4)_{\bar uGu} (27)_\rho\nnb\\
&&(157)_{\rho_c}(142)_{r_K}(^{-15}_{+42})_\kappa(62)_{\Gamma_{K}}(60)_{t_c}(30)_\tau
~{\rm MeV}\nnb\\
&=&(70.7\pm 2.8)~{\rm MeV}~.
\label{eq:mstaumu_inst}
\eea
\nin
  \subsection*{\b Final value of $\hat m_{us}$ and $\overline{m}_{us}$}
  \nin
Our final results are  from Eq. (\ref{eq:ms_ope}) for the SVZ expansion and from the 
combination of the one from 
Eqs. (\ref{eq:msinst}) and (\ref{eq:mstaumu_inst}) from the instanton sum rule. One obtains in units of MeV:
\beq
\hat m^{svz}_{us}=117.4\pm 6.4~,~~~~~~~\hat m^{inst}_{us}=73.3\pm 3.9~,
\label{eq:mus}
\eeq
where the errors on $\hat m^{inst}_{us}$ is the quadratic sum of the one  from the most precise determination and the systematics estimated from the distance of the mean to the central value of  this precise determination. 
The corresponding running masses evaluated at 2 GeV are:
\beq
\overline{m}_{us}^{svz}=101.1\pm 5.5~,~~~~~\overline{m}_{us}^{inst}=63.1\pm 3.4~.
\label{eq:musrun}
\eeq
Using as input the values of $\overline{m}_u$ given in Eq. (\ref{eq:mfinal}), one can deduce:
\beq
\overline{m}_{s}^{svz}=98.5\pm 5.5~,~~~~~\overline{m}_{s}^{inst}=61.5\pm 3.4~.
\label{eq:msrun}
\eeq
Combining this result with the value of $\overline{m}_{ud}$ in Eq. (\ref{eq:mud_final}),
one predicts the scale-independent mass ratios:
\beq
\ga{m_{s}\over m_{ud}}\dr^{svz}=24.9\pm 2.3~,~~~~~\ga{m_{s}\over m_{ud}}\dr^{inst}=25.4\pm 2.2~,
\label{eq:mudsratio}
\eeq
and:
\beq
\ga{m_{s}\over m_{d}}\dr^{svz}=18.7\pm 2.0~,~~~~~\ga{m_{s}\over m_{d}}\dr^{inst}=19.0\pm 2.0~.
\label{eq:mdsratio}
\eeq
These results are summarized in Table \ref{tab:res} where one can remark that unlike the absolute values of the light quark masses, their ratios are almost unaffected by the presence of instanton in the OPE. These results agree within the errors with the previous determinations in \cite{SNB1,SNmass}. One can also compare these results with the recent lattice average for $n_f=2\oplus 1$ flavours \cite{LATT13}:
\beq
 \overline{m}_{s}^{latt}=(93.8\pm 2.4)~{\rm MeV}~, ~~~~~\ga{m_{s}\over m_{ud}}\dr^{latt}=27.44\pm 0.44~,
\eeq
where a good agreement with the ratio and with the value of $\overline{m}_{s}^{svz}$ is observed  while the one of $\overline{m}_{s}^{inst}$ is too low.  
\section{ Lower bounds on $\hat{m}_{uq}$ and $\overline{m}_{uq}$  from Laplace sum rule }
Lower bounds on quark masses have been first derived in \cite{BECCHI,SNRAF} and improved later on in \cite{LELLOUCH} using a finite number of $Q^2$-derivatives of the two-point function. 
Using the second derivative of the two-point function defined in Eq. (\ref{2po}) which is superconvergent:
\beq
\psi_5^{''}(Q^2)=\int_{m_\pi^2}^{\infty}dt{2\over (t+Q^2)^3}{\rm Im}\psi_5(t)~,
\eeq 
retaining the pion pole and using the positivity of the spectral function, one can derive the (linear)  lower bound:
\beq
(\overline{m}_u+\overline{m}_d)(Q^2)\geq {4\pi}\sqrt{2\over 3}{m_\pi^2f_\pi\over Q^2}~,
\lb{eq:boundraf}
\eeq 
at lowest order of PT QCD.
In \cite{SNB2}, the $\alpha_s^3$ corrections to the result to order $\alpha_s^2$ of \cite{LELLOUCH} have been included
leading to the (improved) lower linear bounds evaluated at Q=2 GeV:
\bea
{\overline m}_{ud}&\equiv& {1\over 2}(\overline{m}_u+\overline{m}_d)\geq (3.0\pm 0.5)~{\rm MeV}~,\nnb\\
{\overline m}_{us}&\equiv& (\overline{m}_u+\overline{m}_s)\geq (82.7\pm 13.3)~{\rm MeV}
~,
\label{eq:bound_deriv}
\eea
In \cite{CHET4}, order $\alpha_s^4$ corrections have been added for improving the previous bound on ${\overline m}_{us}$ and lead to a  result consistent with the one in Eq. (\ref{eq:bound_deriv}). However, there is no other arguments for fixing the value of $Q^2$ obtained in Eq. (\ref{eq:boundraf}) apart the convergence  of PT series at which the bound is evaluated. As the $1/Q^2$ fall off of the bound is faster than the $Q^2$ behaviour of the running mass in Eq. (\ref{eq:rgi}) predicted by the RGE, the bound becomes relatively weak at larger $Q^2$-values. 
In the present work, we shall use  the Laplace sum rule ${\cal L}_5^{P}(\tau,\mu)$ (linear constraint) defined in Eq. (\ref{eq:lsr}) together with the optimization procedure used in previous sections 
for extracting an ``optimal  lower bound" on the sum of light quark RGI scale-independent masses $(\hat m_u+\hat m_q)$ which we shall translate later on to bounds on the running quark masses $(\overline{m}_u+\overline{m}_q)$. In so doing, we shall use the positivity of the ``QCD continuum contribution"  by taking $(t_c\to \infty)$ in Eq. (\ref{eq:lsr}) and we shall only consider the meson pole contributions to the spectral function.  We shall also include  (for the first time for the Laplace sum rules) the $\alpha_s^4$ PT corrections for deriving these bounds.
\subsection*{\b Bounds from Laplace sum rules at $\mu=\tau^{-1/2}$}
\begin{figure}[hbt] 
\begin{center}
\centerline {\hspace*{-8.5cm} a) }\vspace{-0.6cm}
{\includegraphics[width=9.5cm  ]{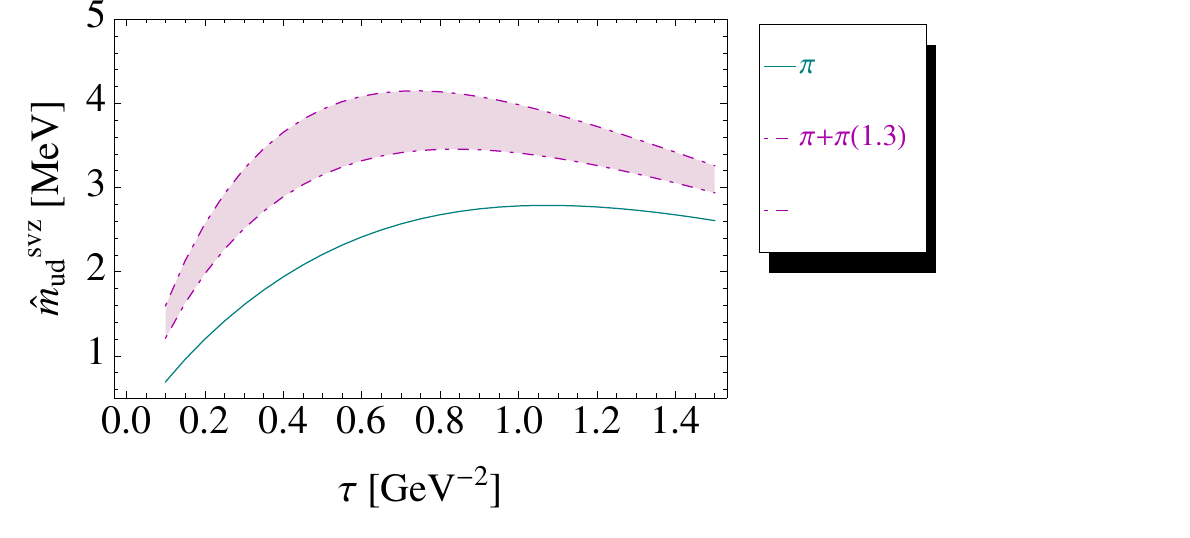}}
\centerline {\hspace*{-8.5cm} b) }\vspace{-0.3cm}
{\includegraphics[width=9.5cm  ]{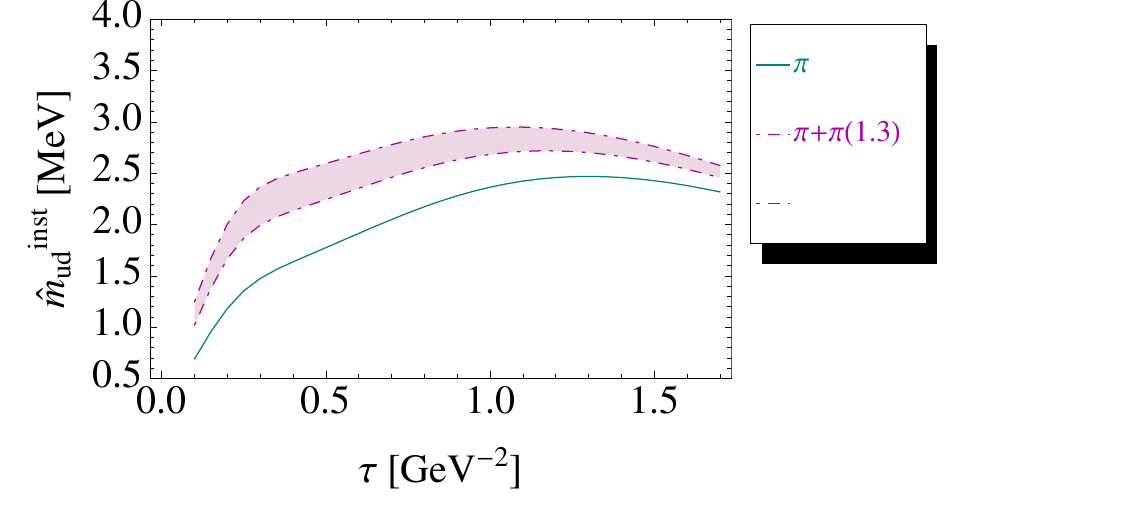}}
\caption{
\scriptsize 
{\bf a)} $\tau$-behaviour  of  the lower bound of $\hat m_{ud}$ from the  sum rule within the SVZ expansion for $\mu=\tau^{-1/2}$. continuous line:  pion contribution only, shaded region: inclusion of the $\pi(1.3)$ ; {\bf b)}  the same as in a) but for the instanton sum rule.}
\label{fig:md_bound}
\end{center}
\end{figure}

\begin{figure}[hbt] 
\begin{center}
\centerline {\hspace*{-8.5cm} a) }\vspace{-0.6cm}
{\includegraphics[width=9.5cm  ]{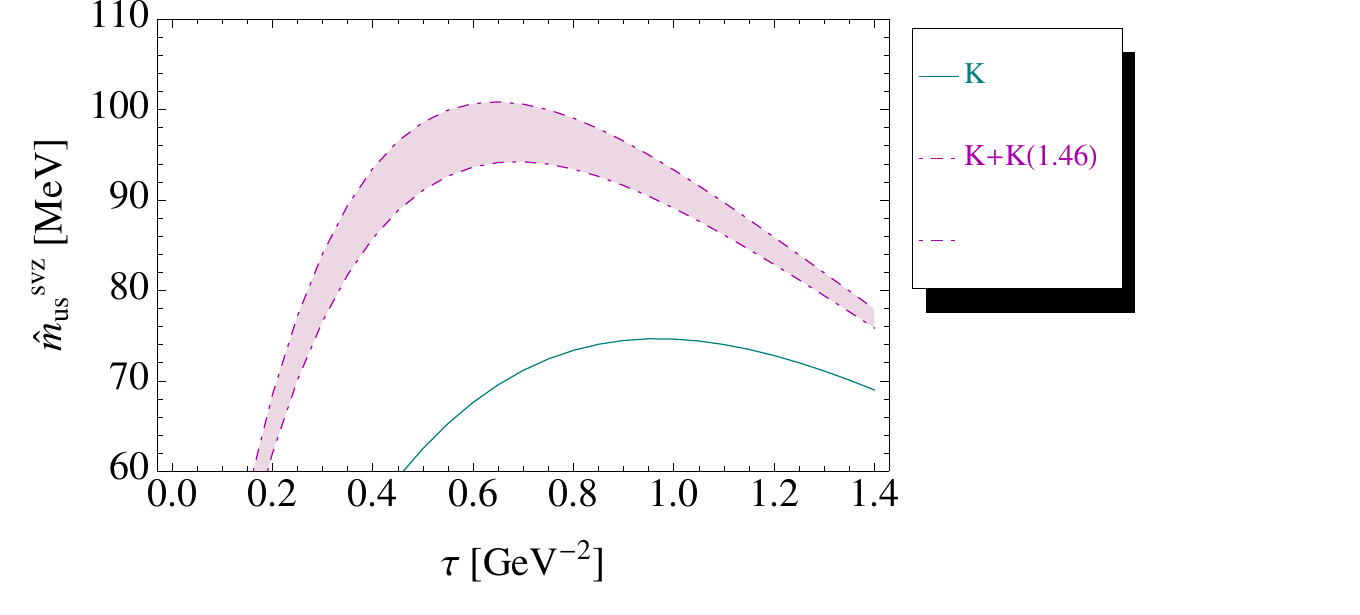}}
\centerline {\hspace*{-8.5cm} b) }\vspace{-0.3cm}
{\includegraphics[width=9.5cm  ]{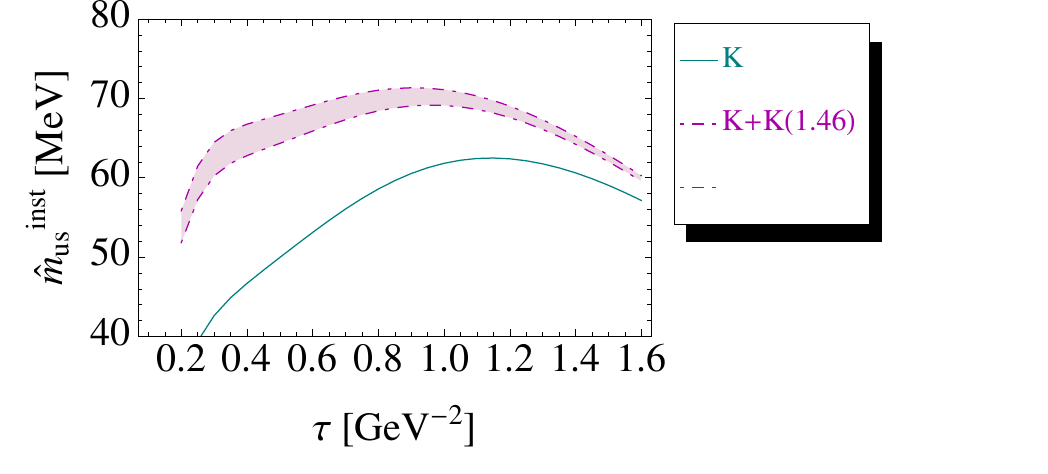}}
\caption{
\scriptsize 
{\bf a)} $\tau$-behaviour  of  the lower bound of $\hat m_{us}$ from the  sum rule within the SVZ expansion for $\mu=\tau^{-1/2}$. continuous line:  kaon contribution only, shaded region: inclusion of the $K(1.46)$ ; {\bf b)}  the same as in a) but for the instanton sum rule.}
\label{fig:ms_bound}
\end{center}
\end{figure}

\nin
We study the lower bounds obtained from ${\cal L}_5^{P}(\tau,\mu)$ sum rule within the SVZ expansion (Fig. \ref{fig:md_bound}a) and the one where the instanton contribution is added into the OPE (Fig. \ref{fig:md_bound}b). We have only retained the pion contribution into the spectral function. Similar curves are obtained in the $s$-quark channel (Fig. \ref{fig:ms_bound}). Among the different bounds associated to $\tau$ shown in these figures, the  most stringent one (hereafter denoted ``optimal bound")  on the quark invariant masses which are scale independent will be extracted at the maximum or / and at the $\tau$-stability region where one has both a good control of the OPE and an optimal contribution of the resonances to the spectral function. 
For $\mu=\tau^{-1/2}$, we obtain\,\footnote{Within the present approximation, the sum rules with an arbitrary $\mu$ do not present a $\tau$-stability and will not be useful here.} in units of MeV:
\bea
\hat m_{ud}^{svz}&\geq&2.79(14)_{\Lambda}(4)_{\lambda^2}(4)_{\bar uu}(8)_{G^2}(0)_{\bar uGu}(9)_\rho\nnb\\
&\geq&(2.79\pm 0.19)~,\nnb\\
\hat m_{us}^{svz}&\geq&74.6(30)_{\Lambda}(10)_{\lambda^2}(7)_{\bar uu}(15)_{G^2}(0)_{\bar uGu}(4)_\rho
(7)_\kappa\nnb\\
&\geq&(74.6\pm 3.7)~,
\eea
and:
\beq
\hat m_{ud}^{inst}\geq2.47(16)_{svz}(0)_{\rho_c}~,~~~~\hat m_{us}^{inst}\geq62.5(28)_{svz}(1)_{\rho_c}~.
\eeq
At the  scale $\tau\approx 1$ GeV$^{-2}$ where these optimal bounds are extracted (maximum in $\tau$),
the NLO, N2LO,N3LO and N4LO PT QCD corrections to these bounds normalized to the LO contributions are respectively: 
\beq
{\rm PT}={\rm LO}\ga 1+0.32+0.22+0.14+0.10\dr~,
\eeq
indicating a reasonnable convergence of the PT series. A duality between the PT series and the tachyonic mass contribution \cite{SNZ} leads to an estimate of about 0.04 of  the uncalculated large order terms contributions. At this scale the d=4 and 6 condensate contributions are respectively 9\% and 8\% of the total PT contributions while the instanton contribution is 28\%. 
One can translate the previous bounds on the RGI masses into the ones for the running masses evaluated at 2 GeV in units of MeV:
\bea
\overline {m}_{ud}^{svz}&\geq&2.41\pm 0.15~,~~~~~~\overline {m}_{us}^{svz}\geq64.3\pm 3.1~,\nnb\\
\overline {m}_{ud}^{inst}&\geq&2.13\pm 0.14~,~~~~~~\overline {m}_{us}^{inst}\geq 53.8\pm 2.4~.
\label{eq:bound_lapl}
\eea
Using the value of the ratio $m_u/m_d$ in Eq. (\ref{eq:ratiomass}), one can deduce from the bound on $\overline{m}_{ud}$ in units of MeV:
\bea
\overline {m}_{u}^{svz}&\geq&1.61\pm 0.10~,~~~~~~\overline {m}_{u}^{inst}\geq 1.42\pm 0.13~,\nnb\\
\overline {m}_{d}^{svz}&\geq&3.21\pm 0.20~,~~~~~~\overline {m}_{d}^{inst}\geq 2.84\pm 0.25~.
\label{eq:boundmu}
\eea
Using the GMOR relation in Eq. (\ref{eq:gmor}), one can translate the previous lower bounds on $\overline {m}_{ud}$ into upper bounds for the running quark condensate evaluated at 2 GeV:
\beq
-\la\bar uu\ra^{svz}\leq (325\pm 7)^3,~~~~~~~~~ -\la\bar uu\ra^{inst}\leq (339\pm 7)^3,
\label{eq:bounduu}
\eeq
and for the spontaneous mass in units of MeV defined in Eq. (\ref{eq:rgi}):
\beq
\mu_u^{svz}\leq 298\pm 6~, ~~~~~~~~~~~~~~~~~\mu_u^{inst}\leq 311\pm6~.
\eeq
Using the value of $\overline {m}_{u}$ in Eq. (\ref{eq:mfinal}), one can deduce, from the bound on $m_{us}$, the ones of running masses at 2 GeV,  in units of MeV:
\beq
\overline {m}_{s}^{svz}\geq61.5\pm 3.1~,~~~~~~~~~~\overline {m}_{s}^{inst}\geq 52.3\pm3.4~.
\label{eq:boundms}
\eeq
Though weaker than the ones in Eq. (\ref{eq:bound_deriv}), these ``optimal bounds" are interesting as previously discussed.   
The  results are summarized in Table \ref{tab:res}. 
\nin
\nin

\subsection*{\b $\pi(1.3)$ and $K(1.46)$ effects on the previous bounds}
\nin
If one includes the contribution of the $\pi(1.3)$ [resp. $K(1.46)$] in the pion [respectively kaon] spectral function, one can improve
the previous bounds. The effect of the $\pi(1.3)$ and $K(1.46)$ are shown respectively  in Fig.~\ref{fig:md_bound} and Fig.~\ref{fig:ms_bound}. The optimal bounds become (in units of MeV):
\bea
\hat m_{ud}^{svz}\vert_{\pi(1.3)}&\geq&3.81(14)_{\Lambda}(6)_{\lambda^2}(4)_{\bar uu}(8)_{G^2}(0)_{\bar uGu}(11)_\rho\nnb\\
&& (8)_{\Gamma_\pi}(38)_{r_\pi}\nnb\\
&\geq&(3.81\pm 0.41)~,\nnb\\
\hat m_{us}^{svz}\vert_{K(1.46)}&\geq&97.8(31)_{\Lambda}(14)_{\lambda^2}(5)_{\bar uu}(13)_{G^2}(1)_{\bar uGu}(11)_\rho\nnb\\
&&
(_{+2}^{-6})_\kappa(1)_{\Gamma_K}(36)_{r_K}\nnb\\
&\geq&(97.8\pm 5.2)~,
\label{eq:bound_res}
\eea
and:
\bea
\hat m_{ud}^{inst}&\geq&2.84(12)_{\pi(1.3)}(17)_{svz}(0)_{\rho_c}~,\nnb\\
\hat m_{us}^{inst}&\geq&70.3(10)_{K(1.46)}(34)_{svz}(4)_{\rho_c}~,
\label{eq:boundinst_res}
\eea
which  can be translated  into the ones for the running masses evaluated at 2 GeV in units of MeV:
\bea
\overline {m}_{ud}^{svz}&\geq&3.28\pm 0.35~,~~~~~~\overline {m}_{us}^{svz}\geq 84.2\pm 4.5~,\nnb\\
\overline {m}_{ud}^{inst}&\geq&2.45\pm 0.18~,~~~~~~\overline {m}_{us}^{inst}\geq60.5\pm 3.1~.
\label{eq:bound_laplpi}
\eea
and:
\bea
\overline {m}_{u}^{svz}&\geq&2.19\pm 0.27~,~~~~~~\overline {m}_{u}^{inst}\geq 1.64\pm 0.16~,\nnb\\
\overline {m}_{d}^{svz}&\geq&4.37\pm 0.54~,~~~~~~\overline {m}_{d}^{inst}\geq 3.27\pm 0.31~.
\label{eq:boundmupi}
\eea
Like previously, the lower bounds on $\overline {m}_{ud}$ can be translated into upper bounds for the running quark condensate evaluated at 2 GeV:
\beq
-\la \overline{\bar uu}\ra^{svz}\leq (294\pm 11)^3,~~~~~~~~~ -\la \overline{\bar uu}\ra^{inst}\leq (324\pm 9)^3,
\label{eq:bounduu2}
\eeq
and for the spontaneous mass in units of MeV defined in Eq. (\ref{eq:rgi}):
\beq
\mu_u^{svz}\leq 267\pm 10~, ~~~~~~~~~~~~~~~~~\mu_u^{inst}\leq 297\pm9~.
\eeq
Using the previous value of $\overline {m}_{u}$ in Eq.  (\ref{eq:mfinal}), one can deduce from Eq (\ref{eq:bound_laplpi}), the bounds, on the running masses evaluated at 2 GeV, in units of MeV:
\beq
\overline {m}_{s}^{svz}\geq81.6\pm 4.5~,~~~~~~~~~~\overline {m}_{s}^{inst}\geq 58.9\pm 3.1~.
\label{eq:boundmsk}
\eeq
The  ``optimal bounds" obtained in Eq.~(\ref{eq:bound_laplpi})  for the quark running masses from the linear sum rules based on the SVZ expansion and including the $\pi(1300)$ (resp. $K(1460)$) are slightly stronger  than the ones given in Eq.~(\ref{eq:bound_deriv}) obtained from finite number of derivatives.  One may consider the present bounds as alternatives to the
ones in the existing literature\,\cite{BECCHI,SNRAF,LELLOUCH,SNB2,CHET4}.
 \section*{Summary and conclusions}
 We have re-estimated the $\pi(1300)$ and $K(1460)$ decay constants using pseudoscalar Laplace sum rules which we have  compared with some existing ones in the literature. We have used these results for improving the determinations of $(m_u+m_q)$: $q\equiv d,s$ from these channels. Our results obtained from the set of parameters in Table \ref{tab:set} are summarized in Table \ref{tab:res}. 
{\scriptsize
\begin{table}[hbt]
\setlength{\tabcolsep}{0.45pc}
 \caption{    
 Values of the set external parameters $(\mu,\tau,t_c)$ obtained at the stability points corresponding to the optimal values of $r_P$ and $m_q$. $\mu$ is  in GeV, $\tau$ in GeV$^{-2}$  and $t_c$ in GeV$^2$. The indices SVZ and inst correspond to the SVZ and SVZ $\oplus$ instanton expansions.}
 
    {\small
\begin{tabular}{lcccc}
&\\
\hline
Observables&$\mu$&$\tau$&$\tau=\mu^{-2}$&  $t_c$   \\
\hline
\\
{\it Pion channel}&\\
$r_\pi^{SVZ}$&$1.4-1.8$&$0.5-0.7$&$1.7-2.1$&$2.0-2.25$\\
$r_\pi^{inst}$&--&$0.8-1.0$&$0.5-0.7$&--\\
$m_{ud}^{SVZ}$&--&$0.7-0.9$&unstable&--\\
$m_{ud}^{inst}$&--&$0.4-0.5$&$0.4-0.7$&--\\
\\
{\it Kaon channel}&\\
$r_K^{SVZ}$&$1.9-2.3$&$0.6-0.9$&unstable&$3.05-3.25$\\
$r_K^{inst}$&--&unstable&$0.5-0.7$&--\\
$m_{us}^{SVZ}$&--&$0.5-0.9$&unstable&$3.2-4.0$\\
$m_{us}^{inst}$&--&$0.35-0.55$&$0.5-0.9$&--\\
\hline
\end{tabular}
}
\label{tab:set}
\end{table}
} 
{\scriptsize
\begin{table}[hbt]
\setlength{\tabcolsep}{0.66pc}
 \caption{
Summary of the main results of this work. For deriving the values and bounds of $\overline{m}_{u,d}$, we have used $m_u/m_d=0.50\pm 0.03$ deduced from the compilation of PDG13 \cite{PDG} .  The estimated value  of $\overline{m}_u$  has been used for an estimate and for giving a bound on $\overline{m}_s$. The running masses $ \overline{m}_q$ evaluated at 2 GeV  are in units of MeV. The bounds for the quark masses are lower bounds while the ones for the $\la \bar uu\ra$ quark condensate are upper bounds.  }
    {\small
\begin{tabular}{llll}
&\\
\hline
Estimates&SVZ& SVZ $\oplus$ instanton&  Eq.   \\
\hline
\\
$f_{\pi'}/f_\pi$&$(2.42\pm 0.43)10^{-2}$ & $(1.77\pm 0.28)10^{-2}$&\ref{eq:rpifinal}\\
$f_{K'}/f_K$&$(23.5\pm 1.6)10^{-2}$ &$(17.6\pm 1.0)10^{-2}$&\ref{eq:rk_ope}, \ref{eq:rk_inst}\\
\\
$\overline{m}_{ud}$&$3.95\pm 0.28$ &$2.42\pm 0.16$&\ref{eq:mud_final}\\
$\overline{m}_{u}$&$2.64\pm 0.28$ &$1.61\pm 0.14$&\ref{eq:mfinal}\\
$\overline{m}_{d}$&$5.27\pm 0.49$ &$3.23\pm 0.29$&\ref{eq:mfinal}\\
\\
$\la \overline{\bar uu}\ra$&$-(276\pm 7)^3$&$-(325\pm 7)^3$&\ref{eq:bounduu}\\
\\
$\overline{m}_{us}$&$101.1\pm 5.5$ &$63.1\pm 3.4$&\ref{eq:musrun}\\
$\overline{m}_{s}$&$98.5\pm 5.5$ &$61.5\pm 3.4$&\ref{eq:msrun}\\
$m_{s}/ m_{ud}$&$24.9\pm 2.3$ &$25.7\pm 2.2$&\ref{eq:mudsratio}\\
$m_{s}/ m_{d}$&$18.7\pm 2.0$ &$19.0\pm2.0$&\ref{eq:mdsratio}\\
\hline
\end{tabular}

}
\setlength{\tabcolsep}{0.1pc}
{\small
\begin{tabular}{llllll}
\\
\\
\hline
Bounds  &\multicolumn{2}{c}{SVZ} 
                 & \multicolumn{2}{c}{SVZ $\oplus$ Instanton}&Eq. \\
\hline
&$\pi$&$\pi~\oplus~\pi(1.3)$&$\pi$&$\pi~\oplus ~\pi(1.3)$&\\
\cline{2-3} \cline{4-5}
$\overline{m}_{ud}$&$2.41\pm 0.15$&$3.28\pm0.35$&$2.13\pm 0.14$&$2.45\pm 0.18$&\ref{eq:bound_lapl}, \ref{eq:bound_laplpi}\\
$\overline{m}_{u}$&$1.61\pm 0.10$ &$2.19\pm 0.27$&$1.45\pm 0.09$&$1.64\pm 0.16$&\ref{eq:boundmu}, \ref{eq:boundmupi}\\
$\overline{m}_{d}$&$3.21\pm 0.20$ &$4.37\pm0.54$&$2.84\pm 0.19$&$3.27\pm 0.31$&\ref{eq:boundmu}, \ref{eq:boundmupi}\\
\\
$\la \bar uu\ra$&$-(325\pm 7)^3$&$-(294\pm11)^3$&$-(339\pm 7)^3$&$-(324\pm9)^3$&\ref{eq:bounduu}, \ref{eq:bounduu2}\\
\\
&$K$&$K~\oplus~K(1.46)$&$K$&$K~\oplus ~K(1.46)$&\\
\cline{2-3} \cline{4-5}
$\overline{m}_{us}$&$64.3\pm 3.1$&$84.2\pm 4.5$&$53.8\pm 2.4$&$60.5\pm 3.1$&\ref{eq:bound_lapl}, \ref{eq:bound_laplpi}\\
$\overline{m}_{s}$&$61.5\pm 3.1$ &$81.6\pm 4.5$&$52.1\pm 2.4$&$58.9\pm 3.1$&\ref{eq:boundms}, \ref{eq:boundmsk}\\
\hline
\end{tabular}
}
\label{tab:res}
\end{table}
} 
The novel features in the present analysis are:
\\
\b  In addition to the usual sum rule evaluated at $\mu=\tau^{-1/2}$ where $\tau$ is the Laplace sum rule variable, we have used an arbitrary subtraction point $\mu$ in the range 1.4-1.8 GeV \cite{BPR} where the best duality between the QCD and experimental sides of the pion sum rules is obtained. Its most precise values have been fixed from a $\mu$-stability criterion [inflexion point (Figs. \ref{fig:rpi}b and \ref{fig:ms}b)  or an (almost) stable plateau (Figs. \ref{fig:rk} and \ref{fig:msinst}) or an extremum (Fig. \ref{fig:rpinst}) depending on the sum rule used] and is given in Table \ref{tab:set}. The sets of $(\tau,t_c)$,  values parameters which optimize the duality between the experimental and QCD sides of each sum rule  from stability criteria are  summarized in Table \ref{tab:set} and come from Figs. \ref{fig:rpi}a, \ref{fig:rpinst}a, \ref{fig:rpimutau}, \ref{fig:mudtau} to \ref{fig:rk}a, \ref{fig:rkmutau}, \ref{fig:rkinst}a, \ref{fig:ms}a,  \ref{fig:msinst}a and \ref{fig:mstaumu_inst}. Their values may  differ for each form of the sum rules analyzed due to the different reorganization of the QCD series and the relative weight of different resonances in the spectral integral for each form of sum rules. In most cases analyzed in this paper, the Laplace sum rules within the SVZ expansion and for arbitrary value of $\mu$ show a large region of plateau stability, while the $\tau^{-1/2}=\mu$ and the one within the SVZ expansion $\oplus$ instanton show only extremal points which in some cases are reached for large values of $\tau$ and induce some additional errors not present in the one within the SVZ expansion and for arbitrary value of $\mu$.  
\\
\b Unlike the well-known case of $\rho$ meson channel, where the continuum threshold $t_c$ can be interpreted to be approximately the value of the 1st radial excitation $\rho'$ meson mass  \cite{SNB1,SNB2}, the situation for the pseudoscalar mesons are quite different due, presumably, to the Goldstone nature of the pion, where the 1st radial excitation $\pi(1300)$  strongly dominates in the estimate of $r_\pi$ while they act almost equally in the determination of $m_{ud}$. Also, in this pseudoscalar channel, there can be a possible negative interference between the $\pi(1300)$ and the second radial excitation $\pi(1800)$ as emphasized by \cite{BPR}, which is not the case of the $\rho$ meson channel. For this reason, it may be misleading to give a physical interpretation of $t_c$ here as its value should be affected by the relative weight  between the contributions of the two different resonances $\pi$ and $\pi(1300)$ and their eventual interferences in the spectral integral as well as the reorganization of the QCD perturbative and non-perturbative series in each sum rules. Here, the alone solid constraint that one can impose is that $t_c$ should be above the mass of the lowest resonances $\pi(1.3)$ and $K(1.46)$ analyzed where $t_c$ is  2 (resp 3) GeV$^2$ for the pion (resp. kaon) channel. 
\\
\b The improved model-independent extraction of the experimentally unknown contribution of the $\pi(1300)$ and $K(1460)$ into the spectral function and the inclusion of finite width corrections. These results agree with the models presented in Fig. \ref{fig:pionexp} and Fig. \ref{fig:kaonexp}. One can also notice that the contributions of the 2nd radial excitation $\pi(1.8)$ and $K(1.8)$ are negligible as shown explicitly in Fig. \ref{fig:rpi}c.
\\
\b An inclusion of the tachyonic gluon mass into the SVZ expansion showing that its effect is relatively small (it decreases $r_\pi$ and $r_K$ by 0.1 and $\hat m_{ud}$ (resp. $\hat m_{us}$) by 0.13 (resp. 3) MeV)  which is reassuring. This negligible effect together with the picture  of duality  \cite{SNZ} between the tachyonic gluon mass contribution   and the sum of uncalculated higher order terms of the QCD PT series indicates that these large order effects are negligible at the scale where we extract the optimal results which can be explicitly checked  by an estimate of the N5LO contribution based on the geometric growth of the PT series. 
\\
\b An explicit study of the Laplace sum rule including instanton contribution which we have considered as an alternative determination of  $(m_u+m_d)$ despite the controversial role of the instanton into the pseudoscalar sum rule. However, the relative small contribution of the $\pi(1300)$ and $K(1460)$ to the spectral function from this analysis (see the comparison in  Fig. \ref{fig:pionexp} and Fig. \ref{fig:kaonexp}) does not (a priori) favour this contribution which consequently induces relative small values of the quark masses compared to the one using the standard SVZ expansion (see Table \ref{tab:res} and \cite{SNB1,PDG}) and recent lattice calculations \cite{LATT13,PDG}.
\\ 
\b One may consider our results as improvements of the existing analytical determinations of $m_{ud}\equiv (m_u+m_q)/2$ : $q\equiv d,s$  from the pseudoscalar Laplace sum rules since the first analysis of~\cite{BECCHI}.  We have not taken the mean value of the two different determinations from SVZ without instanton and from SVZ $\oplus$ instanton due to the controversial instanton role into the pseudoscalar sum rules.  The results using the SVZ expansion without the instanton contribution can be compared  with previous determinations from the (pseudo)scalar sum rules \cite{SNB0,SNB1,SNB2,SNmass,SNmass98,SNRAF,SCAL,TREL,LARIN,PSEU,BPR,PDG,LELLOUCH,DOM13,DOM,MALT,CHETL}, the ones from $e^+e^-$ \cite{SNTAU95,SNmass} and $\tau$-decay  \cite{GAMIZ,SNGTAU5,SNTAU95} data and from nucleon and heavy-light sum rules \cite{SNmass98}
recently reviewed in \cite{SNmass,SNB1}. 
\\ 
\b Our bounds using Laplace sum rules including PT corrections to order $\alpha_s^4$ are new and might be considered as alternatives   of the existing bounds in the literature  \cite{BECCHI,SNRAF,LELLOUCH,SNB2,CHET4}.  
We plan to review these different determinations in a future publication~\cite{QCD14}. 

\end{document}